\documentclass[11pt]{article}
\usepackage{graphicx}
\usepackage{epsfig}
\usepackage{color}

\newcommand{\BABARPubYear}    {06}

\newcommand{\BABARConfNumber} {008}
\newcommand{\SLACPubNumber} {12024}

\input pubboard/babarsym

\long\def\inst#1{\par\nobreak\kern 4pt\nobreak
    {\it #1}\par\vskip 10pt plus 3pt minus 3pt}

\begin{document}
{\pagestyle{empty}

\begin{flushright}
\babar-CONF-\BABARPubYear/\BABARConfNumber \\
SLAC-PUB-\SLACPubNumber \\
July 2006 \\
\end{flushright}

\par\vskip 5cm

\begin{center}
\Large \bf $\Xi_c'$ Production at \babar
\end{center}
\bigskip

\begin{center}
\large The \babar\ Collaboration\\
\mbox{ }\\
\today
\end{center}
\bigskip \bigskip

\begin{center}
\large \bf Abstract
\end{center}
Using 232~\invfb of data collected by the \babar\ detector,
the $\Xi_c^{'+}$ and $\Xi_c^{'0}$ baryons are reconstructed through the decays:
$\Xi_c^{'+} \rightarrow \Xi_c^+ \gamma$ and $\Xi_c^{'0} \rightarrow \Xi_c^0 \gamma$,
where $\Xi_c^+ \rightarrow \Xi^- \pi^+ \pi^+$ and $\Xi_c^0 \rightarrow \Xi^- \pi^+$.
By measuring the efficiency-corrected yields in different intervals of the
center-of-mass momentum, the production rates from $B$ decays and from the
continuum are extracted. For production from $B$ decays, the branching
fractions are found to be
 $\mathcal{B}(B \rightarrow \Xi_c^{'+} X) \times
  \mathcal{B}(\Xi_c^+ \rightarrow \Xi^- \pi^+ \pi^+)
  = [ 1.69 \pm 0.17 \ \mathrm{(exp.)} \pm 0.10 \ \mathrm{(model)} ] \times 10^{-4}$ and
  $\mathcal{B}(B \rightarrow \Xi_c^{'0} X) \times
  \mathcal{B}(\Xi_c^0 \rightarrow \Xi^- \pi^+)
  = [ 0.67 \pm 0.07 \ \mathrm{(exp.)} \pm 0.03 \ \mathrm{(model)} ] \times 10^{-4}$.
For production from the continuum the cross-sections are found to be
  $\sigma(e^+ e^- \rightarrow \Xi_c^{'+} X) \times \mathcal{B}(\Xi_c^+ \rightarrow \Xi^- \pi^+ \pi^+)
  = 141 \pm 24 \ \mathrm{(exp.)} \pm 19 \ \mathrm{(model)} ~\mathrm{fb}$ and
  $\sigma(e^+ e^- \rightarrow \Xi_c^{'0} X) \times \mathcal{B}(\Xi_c^0 \rightarrow \Xi^- \pi^+)
  = 70 \pm 11 \ \mathrm{(exp.)} \pm 6 \ \mathrm{(model)} ~\mathrm{fb}$.
The helicity angle distributions of $\Xi_c'$ decays are studied
and found to be consistent with $J = \frac{1}{2}$.

\vfill
\begin{center}

Submitted to the 33$^{\rm rd}$ International Conference on High-Energy Physics, ICHEP 06,\\
26 July---2 August 2006, Moscow, Russia.

\end{center}

\vspace{1.0cm}
\begin{center}
{\em Stanford Linear Accelerator Center, Stanford University, 
Stanford, CA 94309} \\ \vspace{0.1cm}\hrule\vspace{0.1cm}
Work supported in part by Department of Energy contract DE-AC03-76SF00515.
\end{center}

\newpage
}

\begin{center}
\small

The \babar\ Collaboration,
\bigskip

%
{B.~Aubert,}
{R.~Barate,}
{M.~Bona,}
{D.~Boutigny,}
{F.~Couderc,}
{Y.~Karyotakis,}
{J.~P.~Lees,}
{V.~Poireau,}
{V.~Tisserand,}
{A.~Zghiche}
\inst{Laboratoire de Physique des Particules, IN2P3/CNRS et Universit\'e de Savoie,
 F-74941 Annecy-Le-Vieux, France }
% ...

\end{center}\newpage

\section{INTRODUCTION}
\label{sec:Introduction}
Charmed baryons have a complex and intricate spectroscopy. The states
can be classified in the following categories:
\begin{center}
\begin{tabular}{ccc}
  \hline
  State       & Quark content & Isospin configuration \\
  \hline
  $\Lambda_c$ & $cud$  & Isosinglet \\
  $\Sigma_c$  & $cqq$  & Isotriplet \\
  $\Xi_c$     & $csq$  & Isodoublet \\
  $\Omega_c$  & $css$  & Isosinglet \\
  \hline
\end{tabular}
\end{center}
where $q$ indicates a $u$ or a $d$ quark. There are numerous states
for each of these quark flavor configurations~\cite{bib:maltman_isgur},
several of which have been observed~\cite{ref:pdg}.

In this analysis, we focus on the $\Xi_c$ states, and in
particular on the lowest resonances above the ground state, the $\Xi_c'$.
This is one of three $\Xi_c$ states, listed below, which are not
radially excited and which have zero orbital angular momentum:
\begin{center}
\begin{tabular}{cccc}
  \hline
  \noalign{\vskip1pt}
  State     & Approx. mass (MeV$/c^2$) & Light flavor wavefunction & $J^P$ \\
  \hline
  \noalign{\vskip1pt}
  $\Xi_c$   & 2470 & Antisymmetric & $\frac{1}{2}^+$ \\
  $\Xi_c'$  & 2580 &  Symmetric    & $\frac{1}{2}^+$ \\
  $\Xi_c^*$ & 2645 &  Symmetric    & $\frac{3}{2}^+$ \\
  \noalign{\vskip1pt}
  \hline
\end{tabular}
\end{center}
Note that the $J^P$ of the $\Xi_c'$ and $\Xi_c^*$ have not been
directly measured but are assigned from
the quark model predictions. In a recent study of the decay
$\Xi_c^0 \rightarrow \Omega^- K^+$~\cite{ref:veronique},
the helicity angle of the $\Xi_c^0$ was found to be consistent
with $J = \frac{1}{2}$, though higher spins were
not excluded.

The mass difference between the $\Xi_c'$ and ground state is only
$\Delta m \equiv m(\Xi_c') - m(\Xi_c) \simeq 107$~MeV$/c^2$.
Hence, 
the $\Xi_c'$ is below threshold for a strong decay via $\Xi_c \pi$
or other final states such as $\Lambda_c \overline{K}$ or $D \Lambda$,
and so it can only decay by photon emission, $\Xi_c' \rightarrow \Xi_c \gamma$. 

The $\Xi_c^{'+}$ and $\Xi_c^{'0}$ states were observed by CLEO
in 5.0~\invfb of data~\cite{bib:cleo:xic2575}.
This observation has not yet been confirmed by another experiment.
Charmed baryons can be produced in two ways at $e^+ e^-$ $B$-factories:
from the continuum and in decays of $B$ mesons.
The CLEO measurement was made with requirements that $x_p \geq 0.5$--$0.6$
depending on the decay channel,
where $x_p \equiv p^* / \sqrt{s/4 - m_{\Xi_c'}^2}$ and
$p^*$ is the center-of-mass momentum of the $\Xi_c'$.
These requirements suppressed combinatorial background, but they also 
removed any $\Xi_c'$ production from $B$-decays, retaining only
the continuum production of $\Xi_c'$.

Recent results from \babar\ and Belle indicate that
$B$ decays to charmed baryon pairs occur at a high rate
even when the available phase space is
small~\cite{belle_b_to_lambdac_lambdac_k,belle_b_to_xic_lambdac,babar_xic,babar_recoil_charm,babar_recoil_charm2}.
One possible explanation is that baryon formation is
enhanced when the two baryons are almost at rest in their
center-of-mass frame, and strongly suppressed when their
relative motion is large (see Ref.~\cite{theory_of_b_to_baryons}
and the references therein). This may also explain
threshold enhancements seen in several modes such as
$B^- \rightarrow p \antiproton K^-$ and
$B^- \rightarrow p \antiproton \Kzb$~\cite{belle_b_to_p_p_k,babar_b_to_p_p_k}.
If processes with low energy release are favored,
the production of low-lying charmed baryon resonances such
as $\Xi_c'$ may be substantial. There is currently no
experimental evidence for the production of $\Xi_c'$
in $B$ decays.

\section{THE \babar\ DETECTOR AND DATASET}
\label{sec:babar}
The data used in this analysis were collected with the \babar\ detector
at the SLAC \pep2\ asymmetric energy storage ring. A total of
$(231.9 \pm 3.5)$~\invfb
of data are used, of which 210.3~\invfb were taken on the $\Upsilon(4S)$
resonance ($\sqrt s = 10.58$~GeV) and the remaining 21.6~\invfb
were taken below the \BB threshold ($\sqrt s = 10.54$~GeV). The on-resonance data
contains $(228.3 \pm 2.5) \times 10^6$ \BB pairs.
The \babar\ detector is described elsewhere~\cite{ref:babar}.

Simulated events with $\Xi_c'$ decaying into the desired final
states are generated for the processes
  $e^+e^- \rightarrow \ccbar \rightarrow \Xi_c' X$ and
  $e^+e^- \rightarrow \Upsilon(4S) \rightarrow \BB \rightarrow \Xi_c' X$,
where $X$ represents the rest of the event.
The \textsc{pythia} simulation package~\cite{ref:pythia},
tuned to the global \babar\ data, is used for the $\ccbar$ fragmentation
and for $B$ decays to $\Xi_c'$, and \textsc{geant4}~\cite{ref:geant4} is used
to simulate the detector response.

\section{ANALYSIS METHOD}
\label{sec:Analysis}

\subsection{Overview}

The $\Xi_c^{'+}$ and $\Xi_c^{'0}$ are reconstructed through
the following decays:
\begin{eqnarray*}
  \Xi_c^{'+} &\rightarrow& \Xi_c^+ \gamma \\
  \Xi_c^{'0} &\rightarrow& \Xi_c^0 \gamma,
\end{eqnarray*}
where the daughter $\Xi_c$ baryons are reconstructed as follows:
\begin{eqnarray*}
  \Xi_c^+   &\rightarrow& \Xi^- \pi^+ \pi^+ \\
  \Xi_c^0   &\rightarrow& \Xi^- \pi^+ \\
  \Xi^-     &\rightarrow& \Lambda \pi^- \\
  \Lambda   &\rightarrow& p \pi^-.
\end{eqnarray*}
The measured invariant mass spectra of the $\Xi_c'$ candidates are
corrected for efficiency. From the $\Xi_c'$ yields in different $p^*$ intervals,
the production rates from $B$ decays and from the
continuum are extracted. We also study the helicity angle ($\theta_h$) distribution.

\subsection{Selection and reconstruction of $\Xi_c^+$ and $\Xi_c^0$}

A $\Lambda$ candidate is reconstructed by identifying a proton
with $\mathrm{d}E/\mathrm{d}x$ and
Cherenkov angle measurements~\cite{ref:babar} and 
combining it with an oppositely charged track interpreted
as a $\pi^-$, and fitting the tracks to a common vertex.
The $\Lambda$ candidate is then combined with a negatively charged
track interpreted as a $\pi^-$, and fitted to a common vertex to form a
$\Xi^-$ candidate. For each $\Lambda$ and $\Xi^-$,
the invariant mass is required to be within 3$\sigma$ of
the central reconstructed value, where $\sigma$ is the fitted mass
resolution (approximately 1.0~MeV$/c^2$ for $\Lambda$
and 1.5~MeV$/c^2$ for $\Xi^-$). The invariant mass is then constrained to the
nominal value~\cite{ref:pdg}.
Each resulting $\Xi^-$ candidate
is then combined with one or two positively charged
tracks interpreted as $\pi^+$ to form a $\Xi_c^0$ or $\Xi_c^+$
candidate. No mass constraint is applied to the
$\Xi_c$ candidates.

Since the $\Xi^-$ has a long lifetime ($c \tau = 4.9$~cm),
we improve the signal-to-background ratio by rejecting prompt background.
The displacement vector from the event primary vertex to the $\Xi^-$ decay vertex is
required to be at least 2.5~mm in the plane transverse to the beam direction.
In addition, the scalar product of the displacement vector with the $\Xi^-$
momentum vector is required to be positive, rejecting unphysical candidates.
These criteria were optimized in
a previous analysis~\cite{babar_xic} and were
finalized before the $m(\Xi_c \gamma)$ spectrum was examined in data.

\begin{figure}
  \begin{center}
    \begin{tabular}{cc}
      \epsfig{file=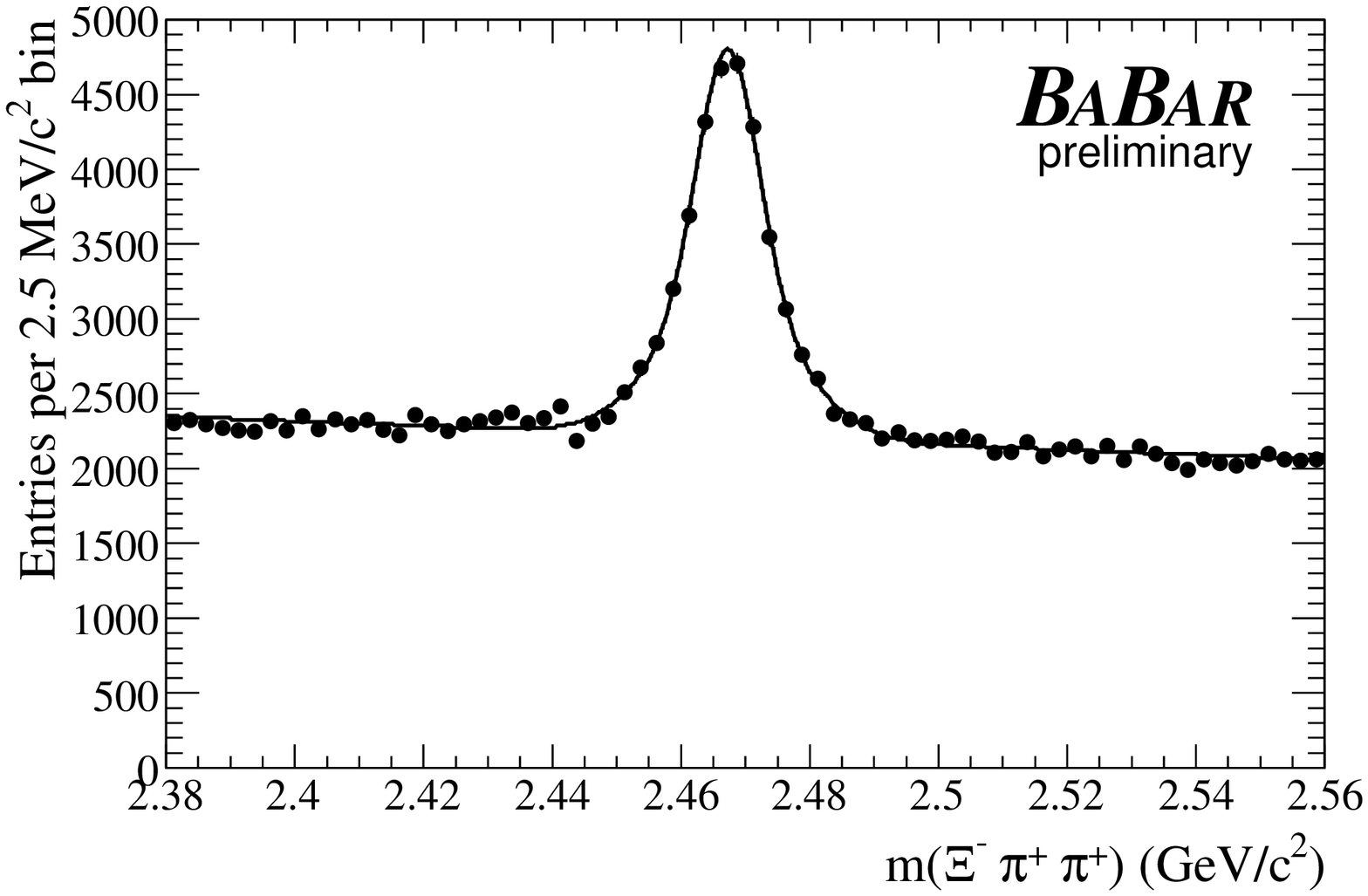, width=0.5\textwidth} &
      \epsfig{file=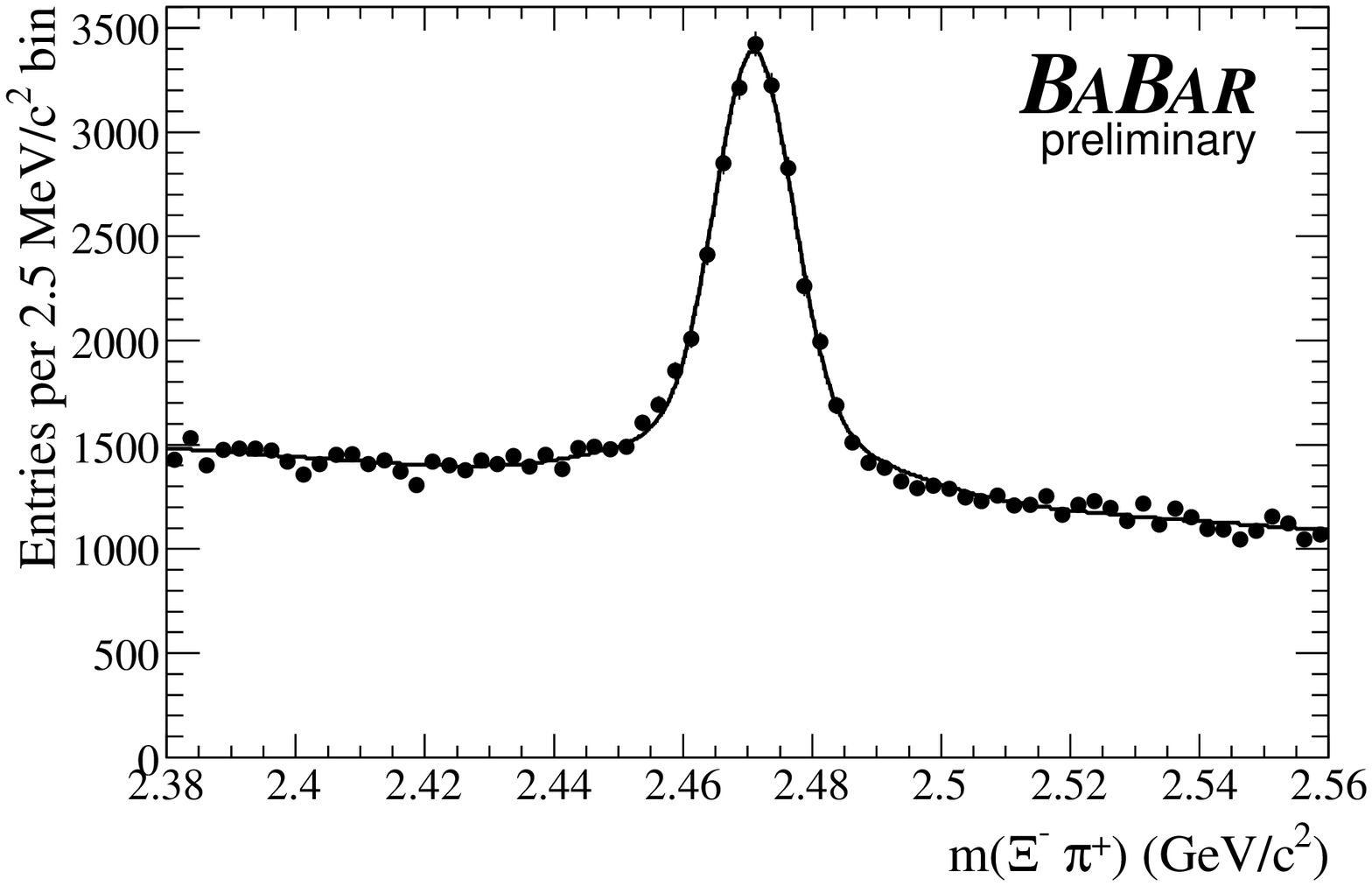, width=0.5\textwidth} 
      \begin{picture}(0.0,0.0)
	\put(-435, 130){\bf(a)}
	\put(-190, 130){\bf(b)}
      \end{picture}
    \end{tabular}
  \end{center}
  \caption[$\Xi_c^+$ and $\Xi_c^0$ mass spectra]
          {Invariant mass spectra of the $\Xi_c$ ground states in 232~\invfb of data.
	   The reconstructed candidates of (a) $\Xi_c^+$ via 
	    $\Xi_c^+ \rightarrow \Xi^- \pi^+ \pi^+$, and 
	    (b) $\Xi_c^0$ via $\Xi_c^0 \rightarrow \Xi^- \pi^+$
             are shown.
	  }
  \label{fig:ground_states}
\end{figure}

The invariant mass distributions for the $\Xi_c$ candidates
satisfying these criteria are shown in
Fig.~\ref{fig:ground_states}. The fitted function (black curve)
is the sum of two Gaussian functions with a common mean for the
signal plus a first-order polynomial for the background.
The half-widths at half-maximum ($\sigma$) of the signal lineshapes
are 7.0~MeV$/c^2$ and 7.5~MeV$/c^2$ for $\Xi_c^+$ and $\Xi_c^0$,
respectively.

\subsection{Reconstruction of $\Xi_c^{'+}$ and $\Xi_c^{'0}$}

\begin{figure}
  \begin{center}
    \begin{tabular}{cc}
      \epsfig{file=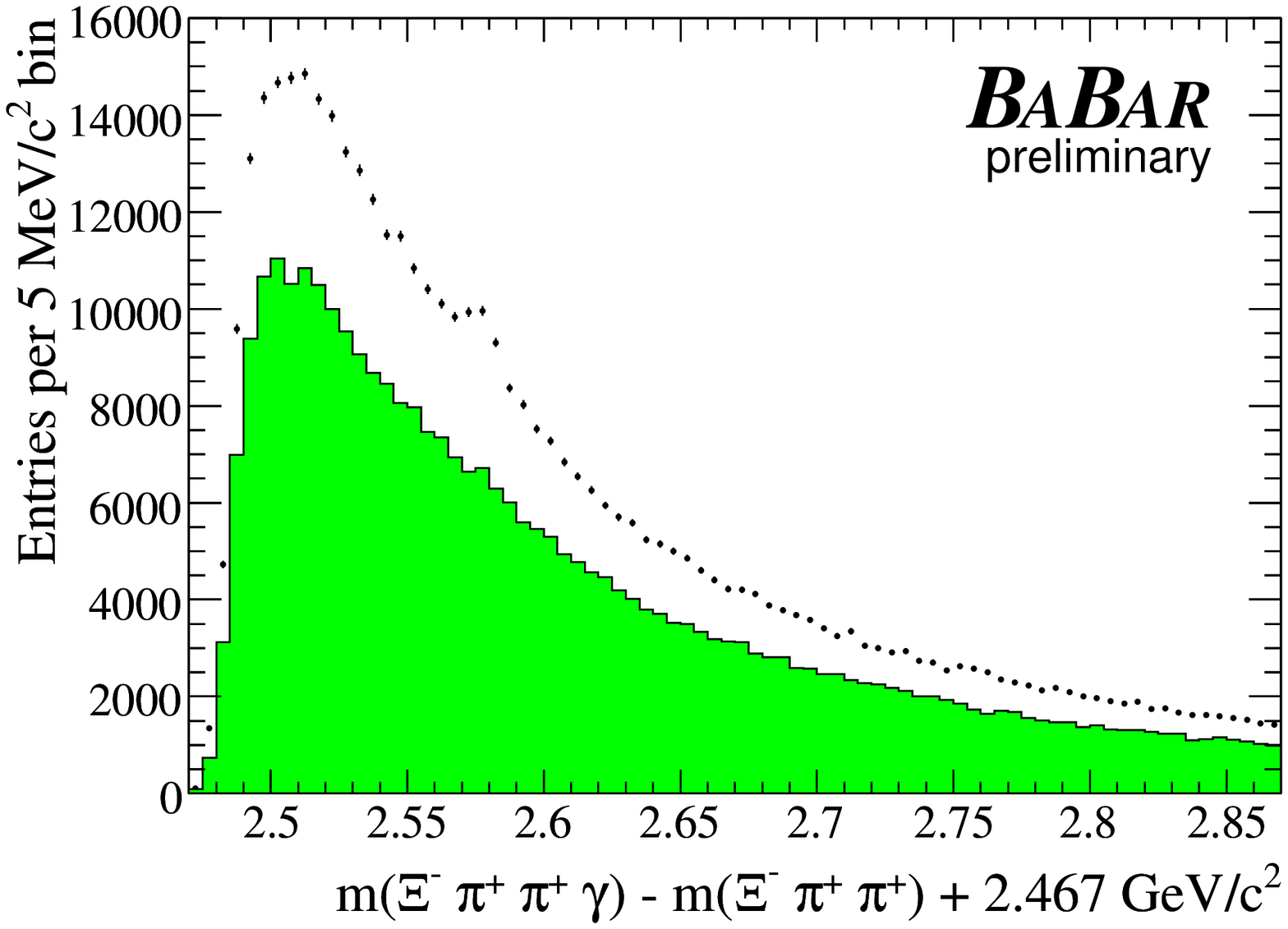, width=0.5\textwidth} &
      \epsfig{file=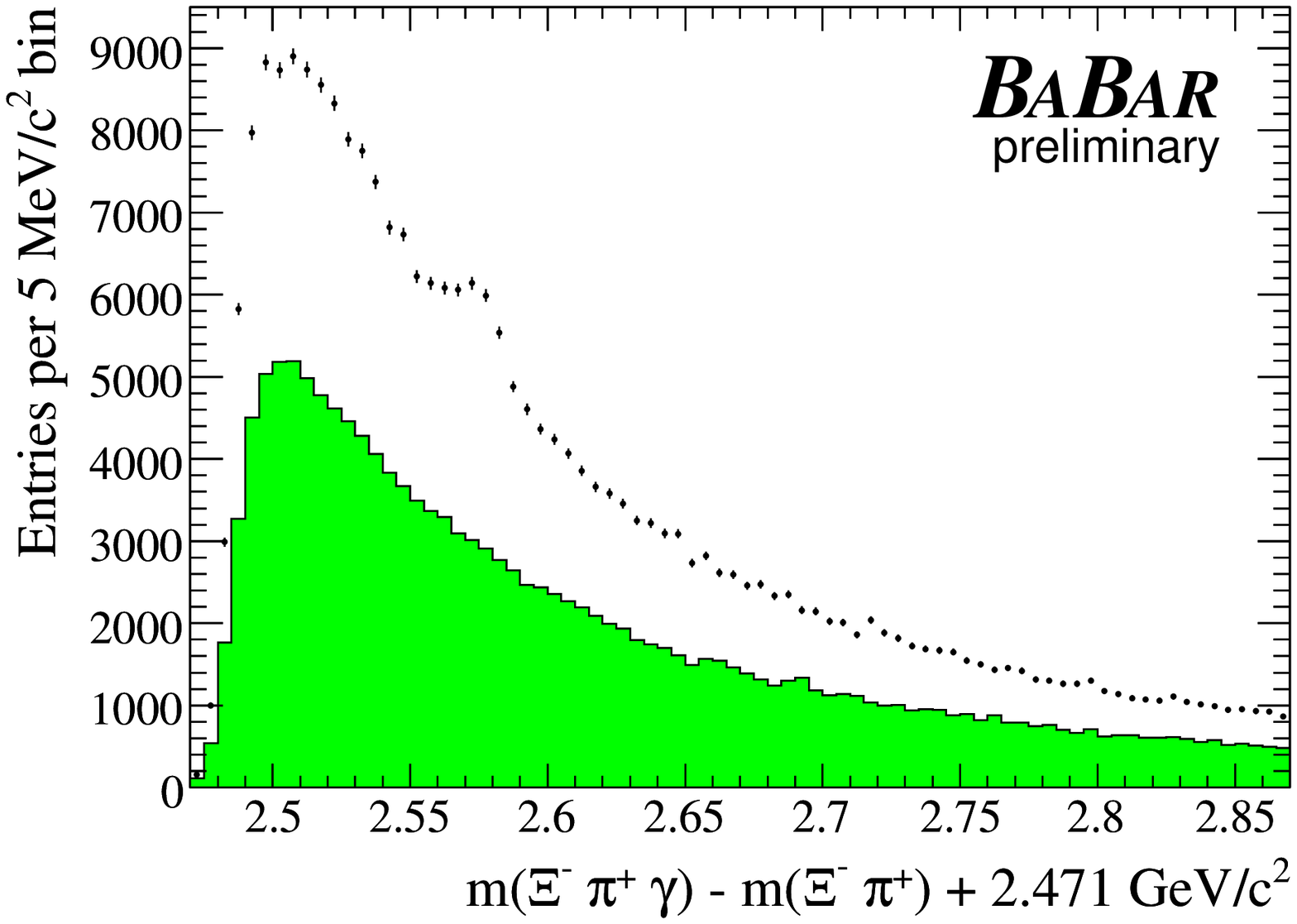, width=0.5\textwidth} \\
      \epsfig{file=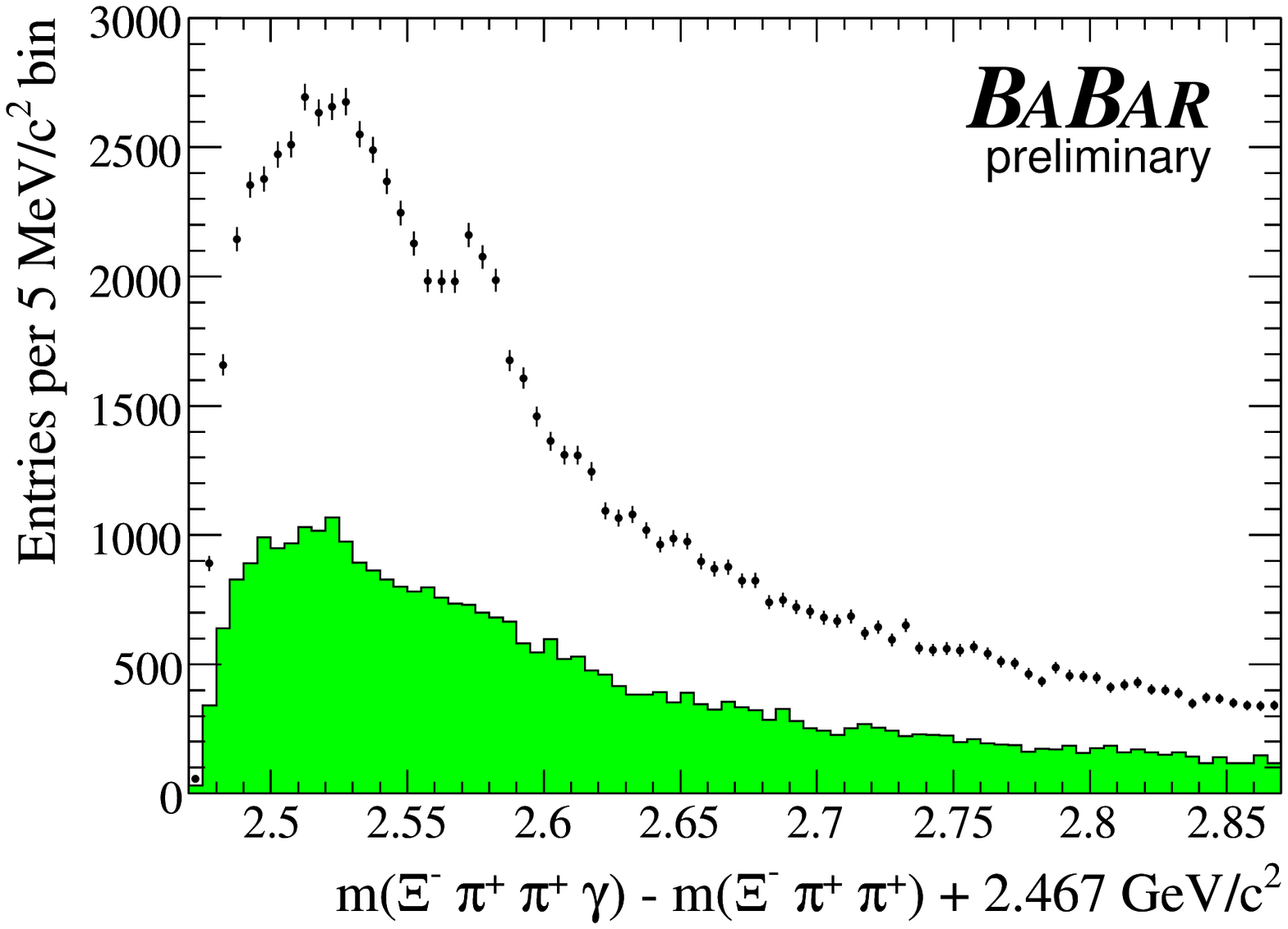, width=0.5\textwidth} &
      \epsfig{file=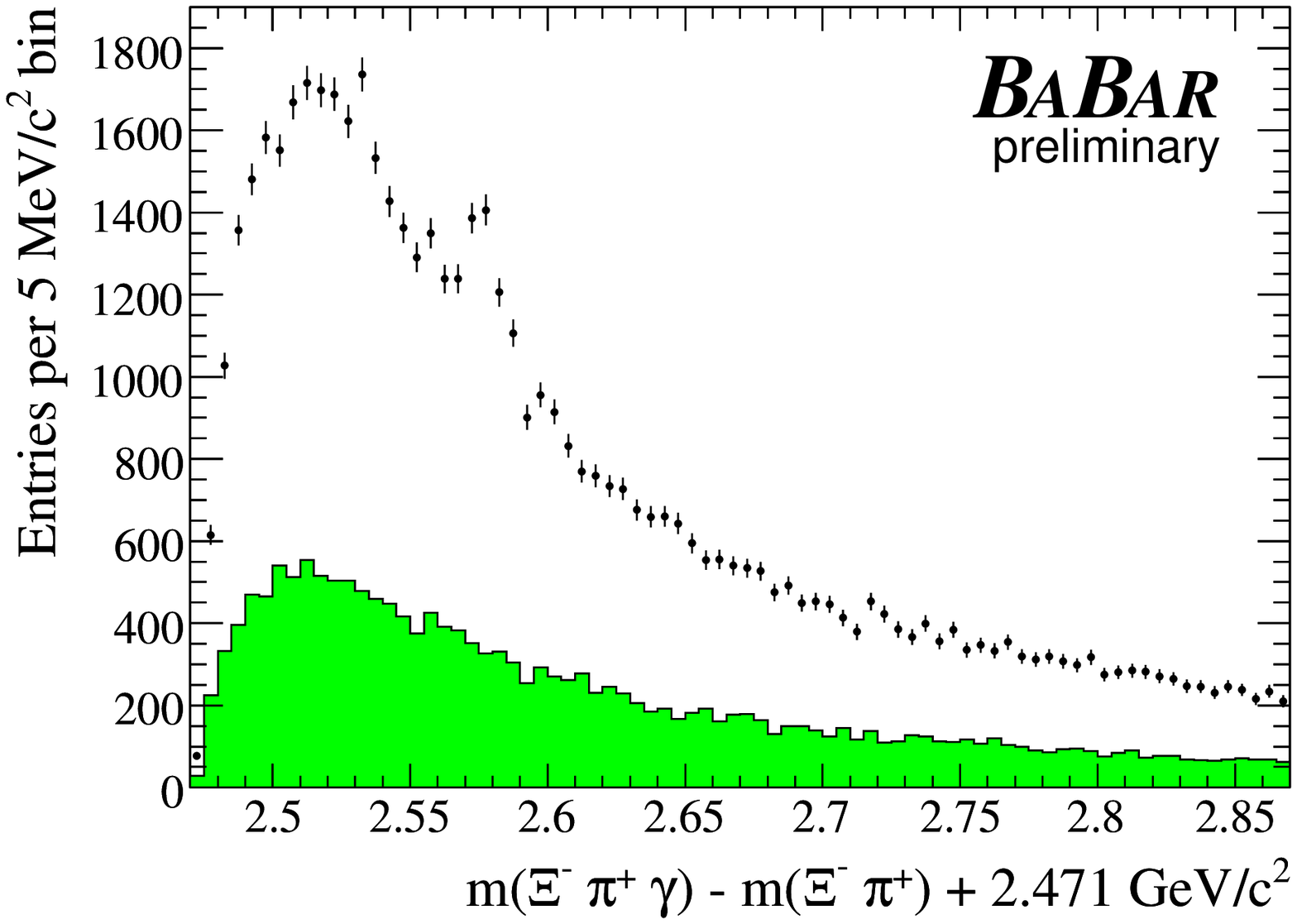, width=0.5\textwidth} \\
      \epsfig{file=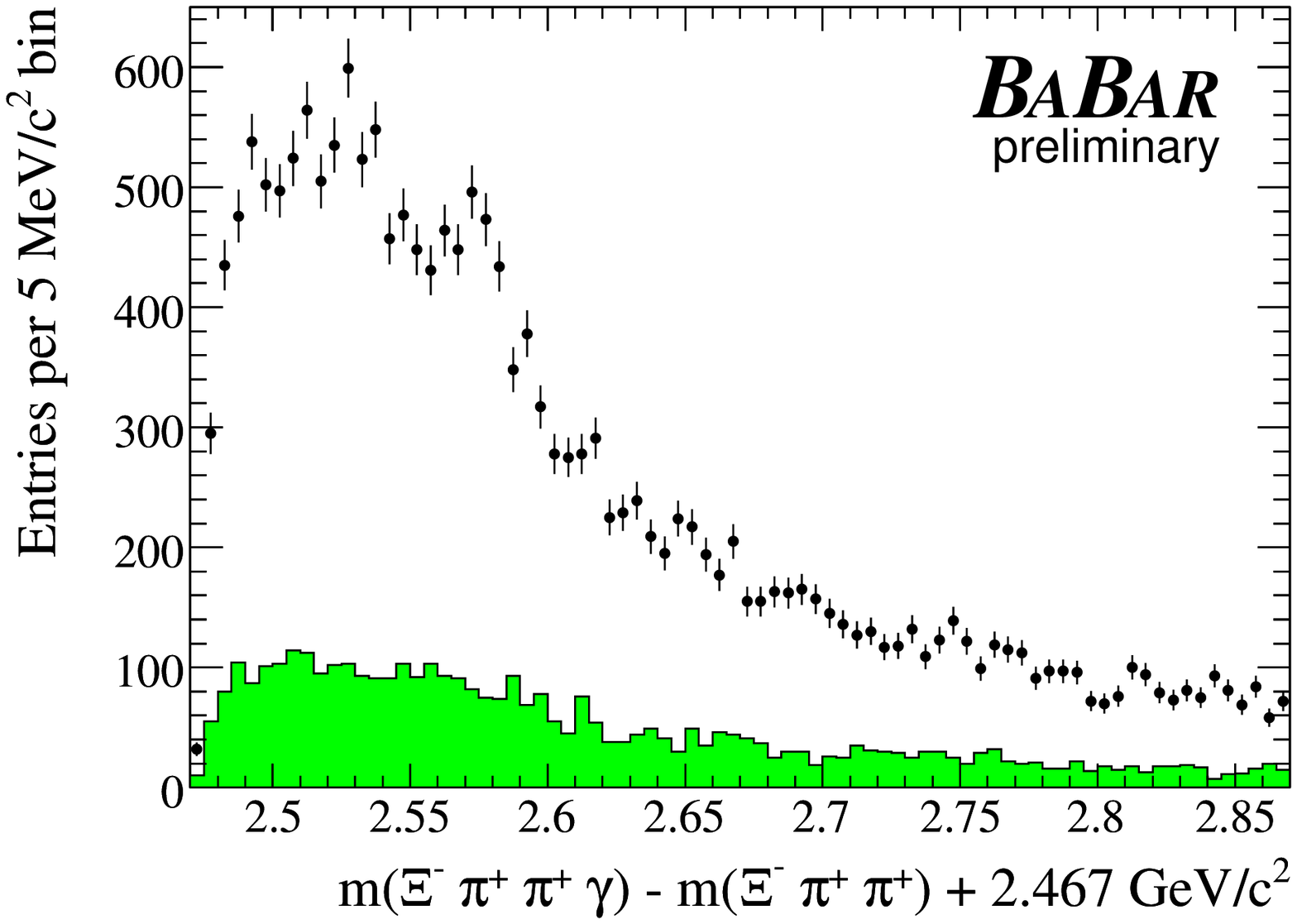, width=0.5\textwidth} &
      \epsfig{file=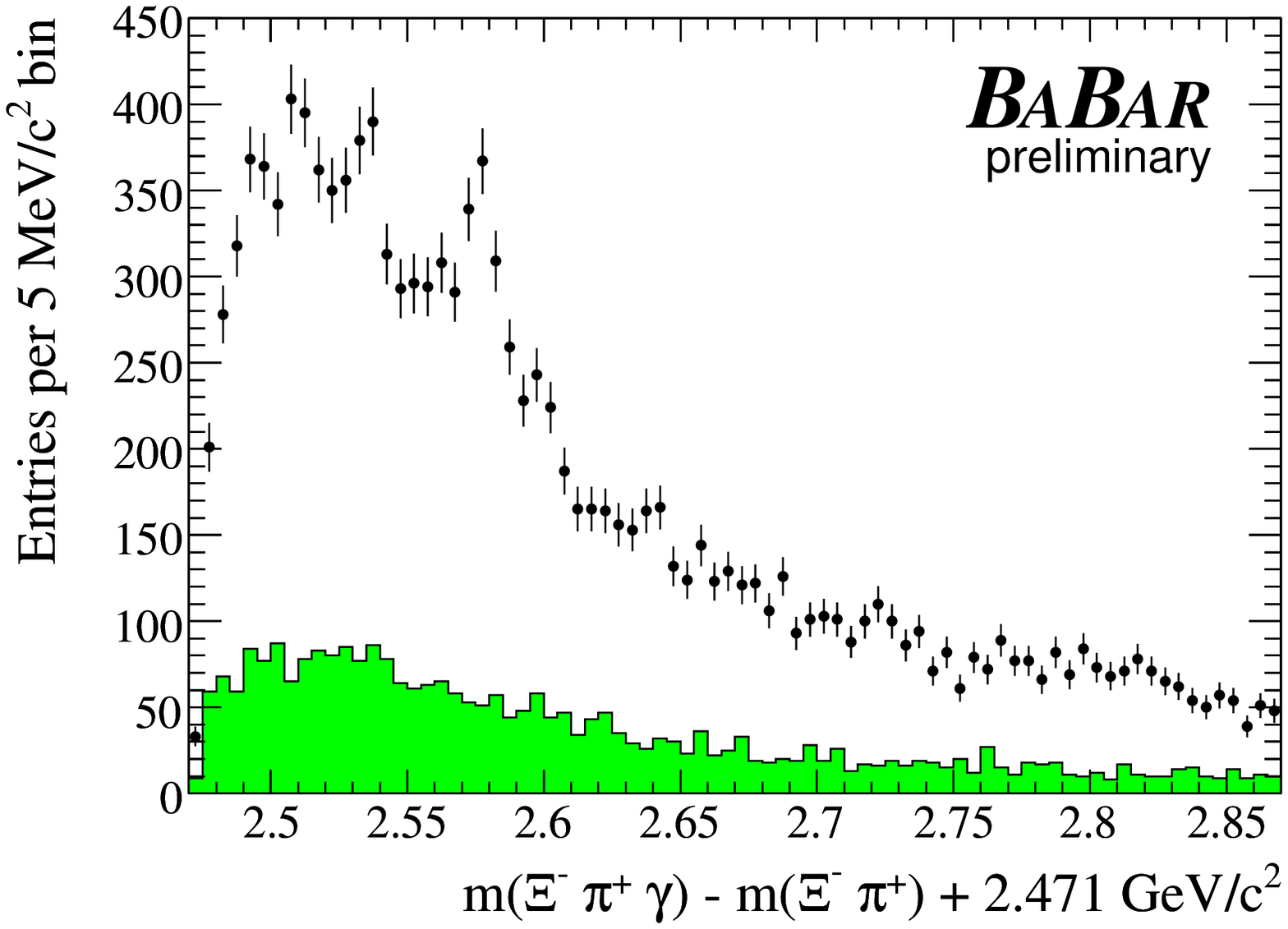, width=0.5\textwidth} 
      
      \begin{picture}(0.0,0.0)
	\put(-300, 460){\bf(a)}
	\put(-50, 460){\bf(b)}
	\put(-300, 285){\bf(c)}
	\put(-50, 285){\bf(d)}
	\put(-300, 110){\bf(e)}
	\put(-50, 110){\bf(f)}
      \end{picture}
    \end{tabular}
  \end{center}
  \caption[Raw $\Xi_c \gamma$ invariant mass spectrum]
          {The $\Xi_c \gamma$ invariant mass spectra, shown with the following $p^*$ requirements:
            0.0~GeV$/c$ (a,b),
            2.5~GeV$/c$ (c,d),
            3.5~GeV$/c$ (e,f).
	    The left column shows $\Xi_c^+ \gamma$ and the right column shows $\Xi_c^0 \gamma$.
	    The shaded histograms are taken from the $\Xi_c$ mass sidebands
	    ($5\sigma$--$8\sigma$ from the central value, where $\sigma$ is the
	    $\Xi_c$ mass resolution), 
	    and the solid points are from the $\Xi_c$ signal region
	    (within $3\sigma$ of the central value).
          }
          \label{fig:spectra:raw_XicGamma}
\end{figure}

Clusters of energy in the electromagnetic calorimeter
are identified. The cluster must spread over at least two crystals.
Clusters which lie along the trajectory of a charged track in
the event are eliminated. The remaining clusters define photon candidates.
The energy of the photon candidate must be at least 30~MeV
and the lateral moment (defined in Ref.~\cite{babar_physics_book})
must be less than 0.8.
The $\Xi_c$ candidates shown in Fig.~\ref{fig:ground_states}
are then combined with each of the photon candidates
to form $\Xi_c'$ candidates. The mass difference $\Delta m$
is then computed:
\begin{displaymath}
  \Delta m = \left\{
  \begin{array}{ll}
    m(\Xi^- \pi^+ \pi^+ \gamma) - m(\Xi^- \pi^+ \pi^+) & \mathrm{for}~\Xi_c^{'+} \\
    m(\Xi^- \pi^+       \gamma) - m(\Xi^- \pi^+      ) & \mathrm{for}~\Xi_c^{'0}
  \end{array}
  \right.
\end{displaymath}
Since $\Delta m$ and $m(\Xi^- \pi^+ [\pi^+])$ are essentially uncorrelated
and the mass difference between $\Xi_c'$ and $\Xi_c$ is small,
this method gives good mass resolution and avoids the need to
apply a mass-constraint to the $\Xi_c$ candidates.
To make the spectra easier to interpret, we plot the mass difference
$\Delta m$ plus a constant offset $m_{\mathrm{offset}}$ which is approximately
equal to the nominal ground state
mass; offsets of 2.467~GeV$/c^2$ and 2.471~GeV$/c^2$ are used
for the $\Xi_c^+$ and $\Xi_c^0$ states, respectively.
The spectra thus obtained are shown for $\Xi_c^+$ and
$\Xi_c^0$ in Fig.~\ref{fig:spectra:raw_XicGamma}.
For each mode, the mass spectrum is shown for 
  $p^* > 0.0$~GeV$/c$ (upper), 
  $p^* > 2.5$~GeV$/c$ (middle), and 
  $p^* > 3.5$~GeV$/c$ (lower).
Clear $\Xi_c'$ signals are observed at a mass of 
approximately 2.58~GeV$/c^2$.

\begin{figure}
  \begin{center}
    \epsfig{file=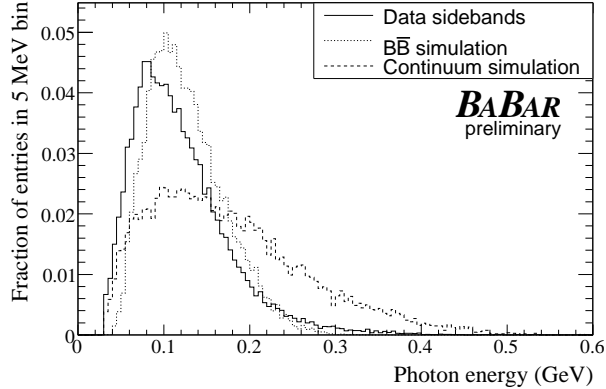, width=0.5\textwidth}
  \end{center}
  \caption[Photon energy distributions for signal and background]
	  {Photon energy distributions. The dotted and dashed histograms
	    show photons from simulated $\Xi_c'$ decays where the
	    $\Xi_c'$ is produced in $B$ decays or from the
	    continuum, respectively. The solid histogram shows
	    photons from $\Xi_c'$ candidates in the data
	    with $m(\Xi^- \pi^+ [\pi^+])$ in the $\Xi_c$ mass sidebands
	    and $\Delta m$ in the signal region
	    ($0.09 < \Delta m < 0.12$~GeV$/c^2$).
	    
	  }
	    \label{fig:photon_selection}
\end{figure}

\begin{figure}
  \begin{center}
    \begin{tabular}{cc}
      \epsfig{file=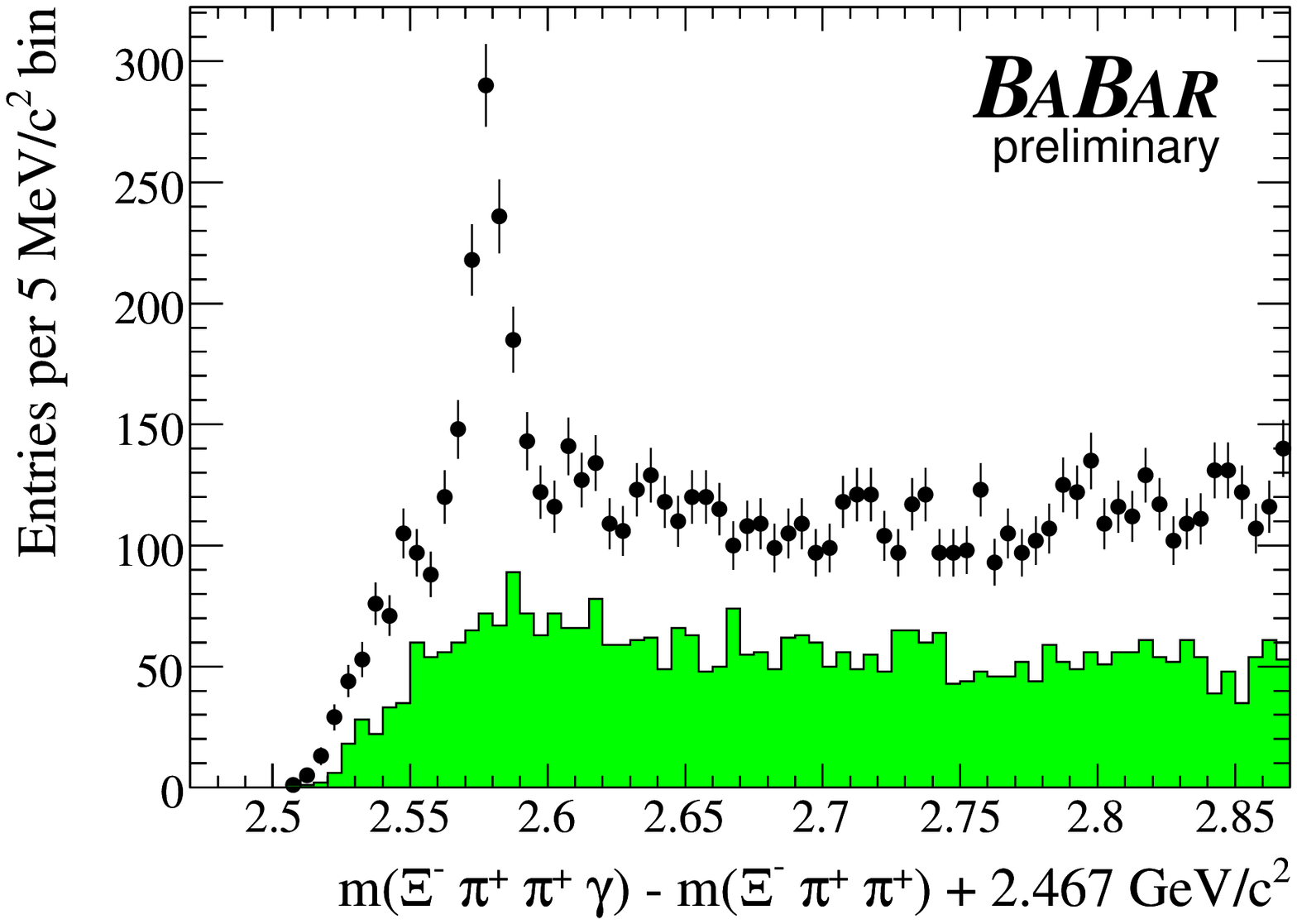, width=0.48\textwidth} & 
      \epsfig{file=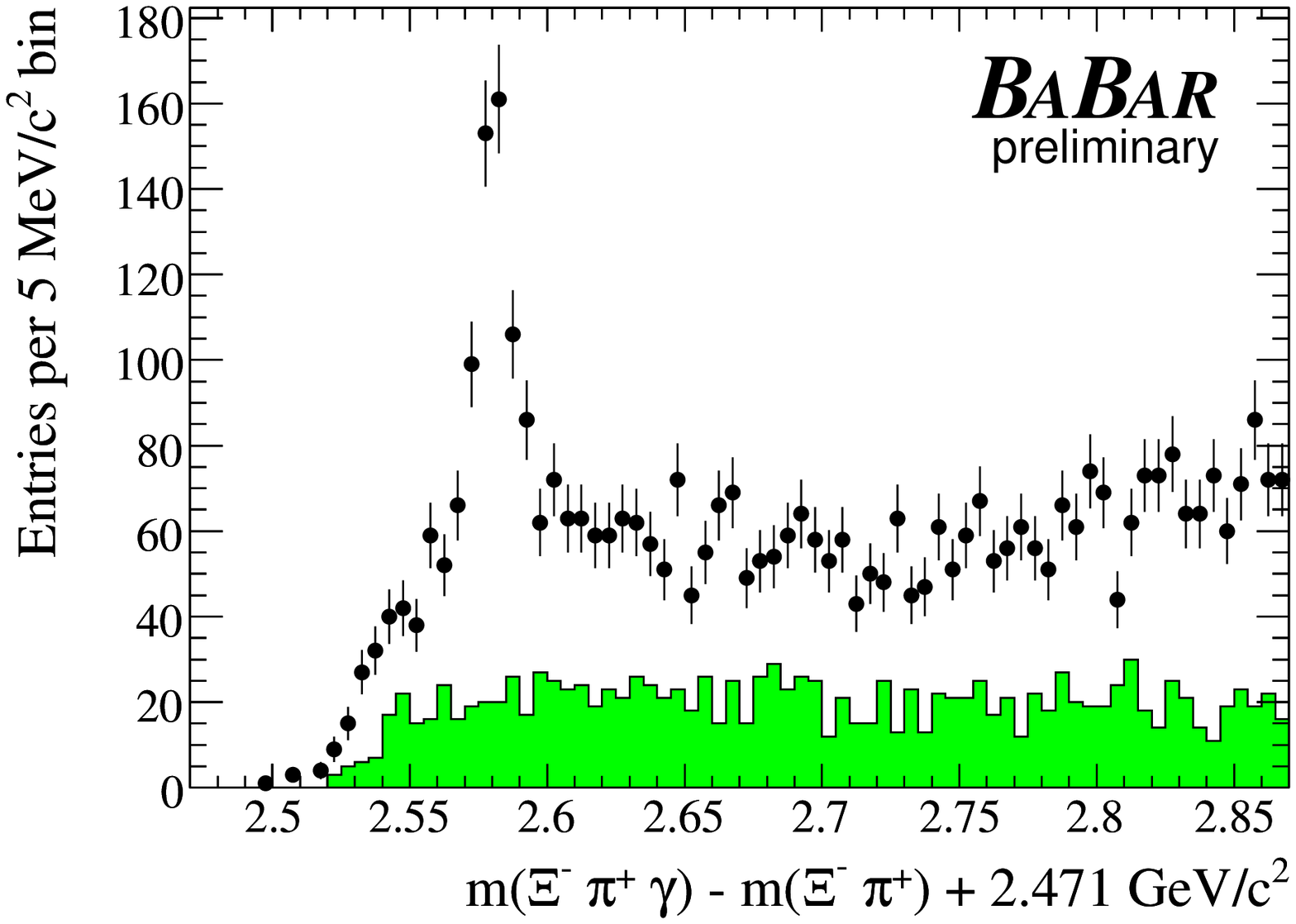, width=0.48\textwidth} 
      \begin{picture}(0.0,0.0)
	\put(-415, 135){\bf(a)}
	\put(-180, 135){\bf(b)}
      \end{picture}
    \end{tabular}
  \end{center}
  \caption[$\Xi_c'$ mass spectra with restrictive photon selection]
	  {
	    The $\Xi_c \gamma$ invariant mass spectra, requiring 
	    that $p^* > 2.5$~GeV$/c$,
	    that the photon energy be above 200~MeV, 
	    that the shower contain at least two crystals, 
	    that the lateral moment be less than 0.6, and
	    that the shower be well-contained (shower energy within two
	    cells of the maximum be at least 90\% of the shower energy
	    within one cell of the maximum). 
	    Plot~(a) shows $\Xi_c^+ \gamma$ and
	    plot~(b) shows $\Xi_c^0 \gamma$.
	    The mass windows are narrower than for
	    Fig.~\ref{fig:spectra:raw_XicGamma}:
	    the shaded histograms are taken from the $\Xi_c$ mass sidebands
	    ($5\sigma$--$7\sigma$ from the central value)
	    and the solid points are from the $\Xi_c$ signal region
	    (within 2$\sigma$ of the central value).
	  }
  \label{fig:photon_selection_spectra}
\end{figure}

The photon selection criteria are quite loose, especially in comparison to the
previous CLEO analysis which imposed a minimum photon energy of
100~MeV in addition to requirements on the lateral shower profile
and a veto of photons from $\pi^0$ candidates. The reason for the
different selection strategies is evident from a comparison of the
energy spectra of photons from $\Xi_c'$ produced in the continuum,
$\Xi_c'$ produced in $B$ decays, and background
(Fig.~\ref{fig:photon_selection}):
the CLEO study excluded $\Xi_c'$ from $B$ decays and was optimized
for sensitivity to continuum production
of $\Xi_c'$ where the photon energy is a powerful discriminant between
signal and background, whereas the selection criteria for
this analysis were chosen to retain $\Xi_c'$ from $B$ decays with
high efficiency. For illustrative purposes, the mass spectra of
$\Xi_c'$ candidates with tighter selection requirements
are shown in Fig.~\ref{fig:photon_selection_spectra}.

\subsection{Contributions to the invariant mass spectra}
\label{sec:background}

\begin{figure}
  \begin{center}
    \begin{tabular}{cc}
      \epsfig{file=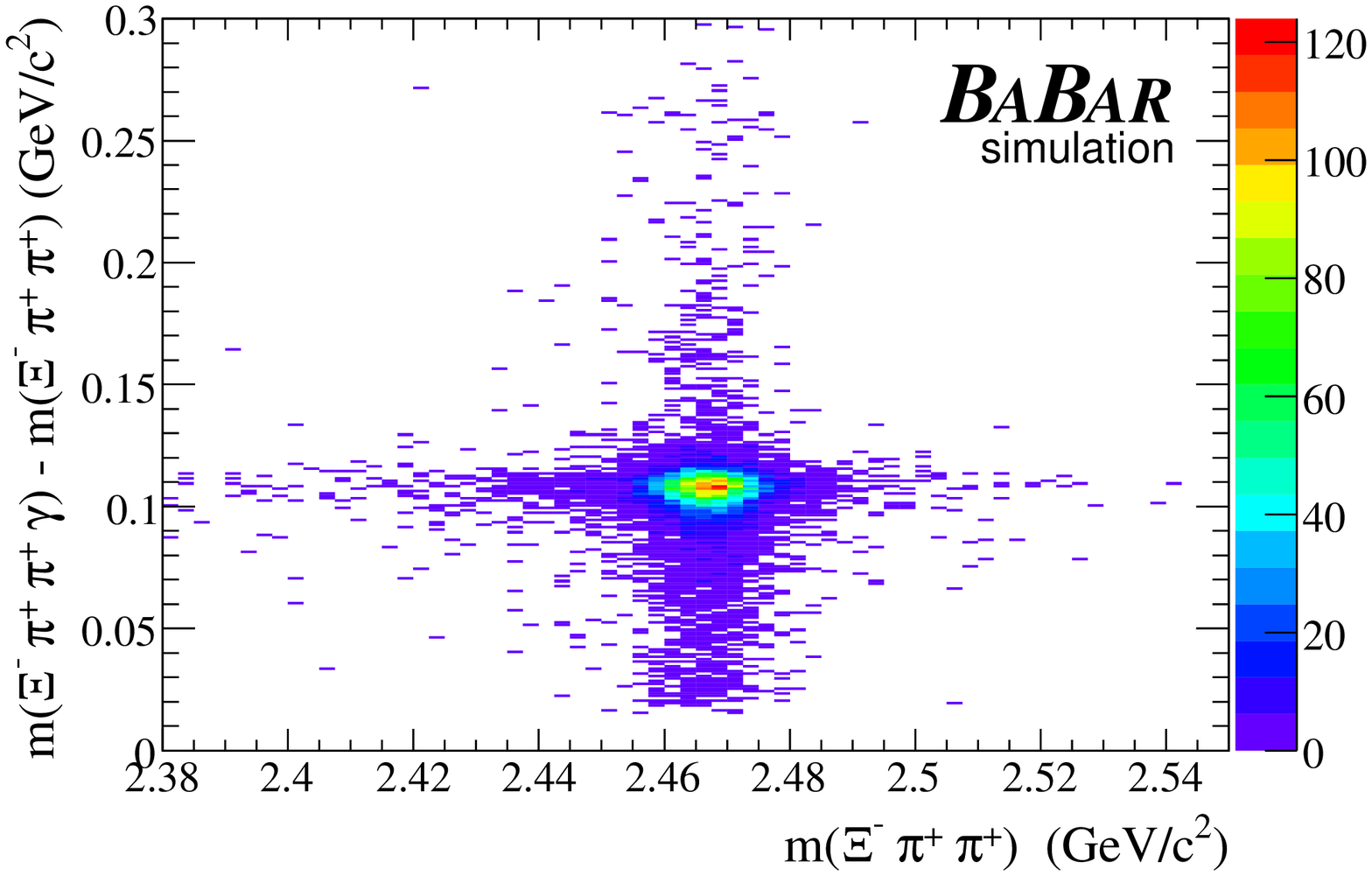, width=0.5\textwidth} &
      \epsfig{file=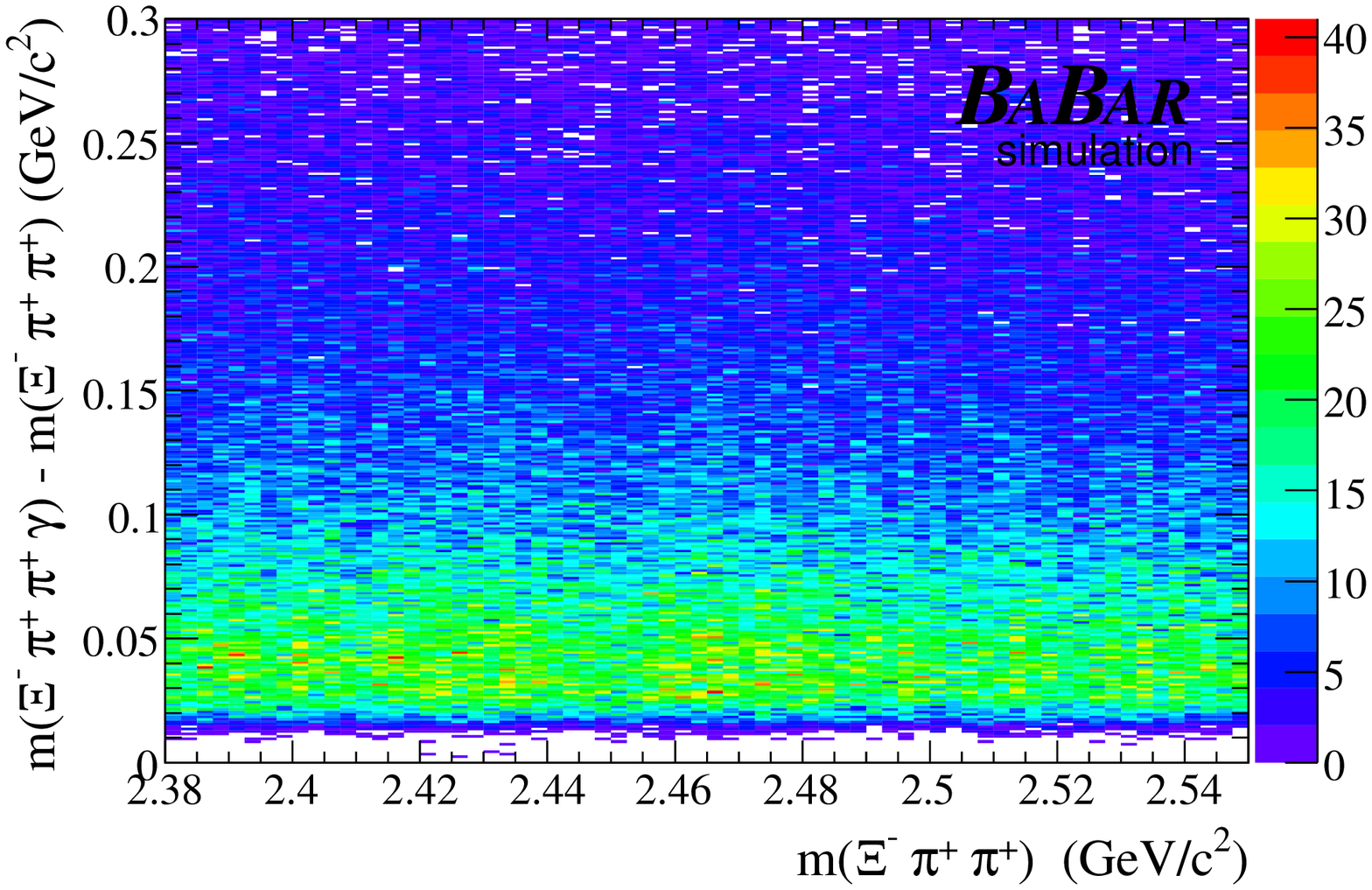, width=0.5\textwidth} \\
      \epsfig{file=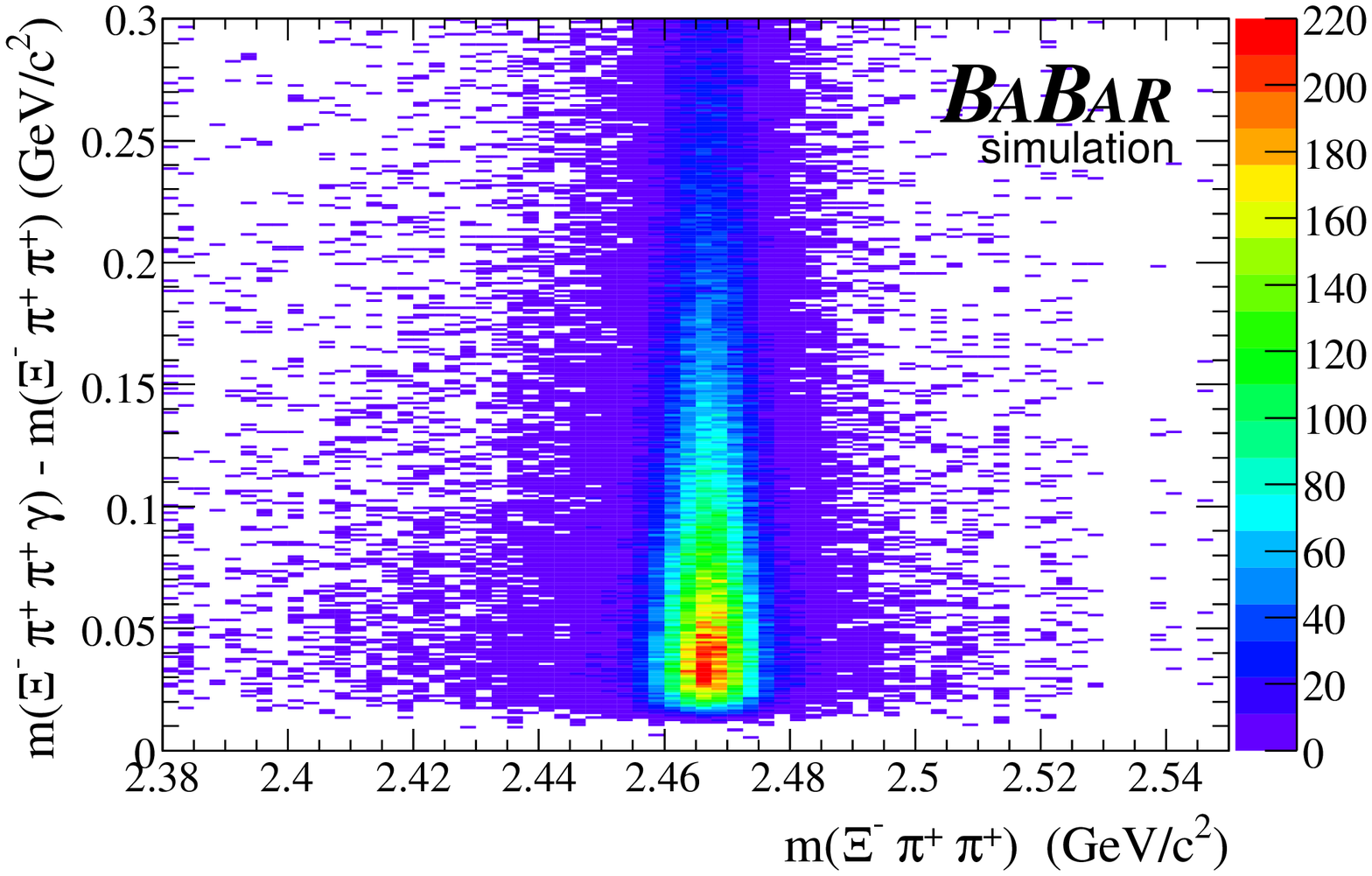, width=0.5\textwidth} &
      \epsfig{file=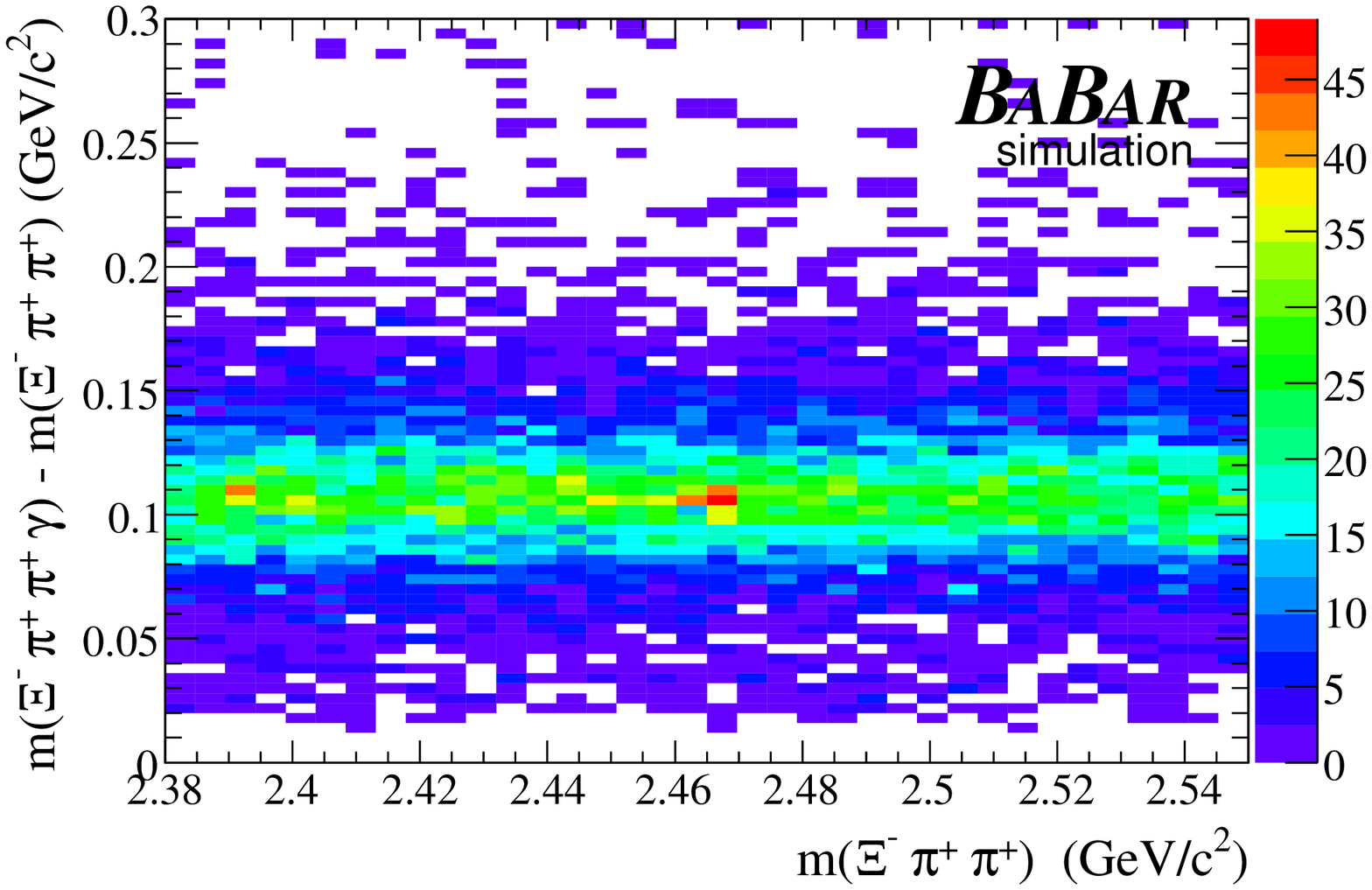, width=0.5\textwidth}
      \begin{picture}(0.0,0.0)
	\put(-450, 300){\bf(a)}
	\put(-200, 300){\bf(b)}
	\put(-450, 140){\bf(c)}
	\put(-200, 140){\bf(d)}
      \end{picture}
    \end{tabular}
  \end{center}
  \caption[Illustrations of the signal and background contributions]
	  {Illustrations of the signal and background contributions for $\Xi_c^0 \gamma$ in simulated continuum events.
	    Plot~(a) shows category 1, correctly reconstructed signal.
	    Plot~(b) shows category 2, combinatoric background.
	    Plot~(c) shows category 3, background with real $\Xi_c$.
	    Plot~(d) shows category 4, where a photon from a real $\Xi_c'$
	    decay is combined with an incorrectly reconstructed $\Xi_c$.
	  }
	  \label{fig:explain_categories}
\end{figure}

Four principal categories of events contribute to the $\Xi_c'$ candidate distributions:
\begin{enumerate}
  \item Signal $\Xi_c' \rightarrow \Xi_c \gamma$ decays
    which peak in both $m(\Xi^- \pi^+ [\pi^+])$ and $\Delta m$.
    This is shown in Fig.~\ref{fig:explain_categories}~(a).
  \item Combinatoric background which does not peak in
    either $m(\Xi^- \pi^+ [\pi^+])$ or $\Delta m$.
    This is shown in Fig.~\ref{fig:explain_categories}~(b).
  \item Background where a real $\Xi_c$ is combined with
    an unrelated photon candidate.
    This peaks in $m(\Xi^- \pi^+ [\pi^+])$ but not in $\Delta m$.
    This is shown in Fig.~\ref{fig:explain_categories}~(c).
  \item Background contribution from events where a real 
    $\Xi_c' \rightarrow \Xi_c \gamma$ decay occurs and the
    correct photon is found but the $\Xi_c$ is partially
    mis-reconstructed. 
    This is shown in Fig.~\ref{fig:explain_categories}~(d).
    This category generally does not peak in $m(\Xi^- \pi^+ [\pi^+])$
    but peaks in $\Delta m$ (provided the momentum
    of the fake $\Xi_c$ candidate is close to the real
    $\Xi_c$ momentum).
\end{enumerate}
Categories 2 and 3 do not peak in $\Delta m$, so we
describe them with a smooth polynomial function. The 
$\Delta m$ distribution of the fourth category is almost
indistinguishable from the signal distribution. 

One further possible contribution to the mass spectrum was considered:
feed-down from the decay $\Xi_c^* \rightarrow \Xi_c \pi^0$
where only one of the two photons produced in the $\pi^0$ decay is used,
leaving the same $\Xi_c \gamma$ final state as for a $\Xi_c'$ decay.
A study of this process with a simple kinematic simulation indicates that
it would produce a very broad, non-peaking structure in the $\Xi_c'$
mass spectrum, and therefore falls into the third category
($\Xi_c$ background not peaking in $\Delta m$) already discussed.

\subsection{Fitting procedure}
\label{sec:fit}

The data are divided into ten $p^*$ intervals of width 0.5~GeV$/c$ from 0.0
to 5.0 \gevc. For each $p^*$ interval, the $\Delta m$ distributions
are fitted with the combination of a signal lineshape extracted from the
simulated signal events and a second-order
polynomial function to describe the background. The signal lineshape
is parameterized as the sum of three Gaussian functions, parameters
of which are determined from a fit to a
high-statistics sample of simulated signal events in the corresponding $p^*$
range. The lineshape is described in more detail in
Section~\ref{appendix:lineshape:gaussian}
of Appendix~\ref{appendix:lineshape}.
During this fit all nine parameters of the triple Gaussian function
are allowed to vary independently within fixed ranges.\footnote{
  Fewer signal events were generated at the extremes of the $p^*$
  spectrum. In cases where fewer than 1,000 signal events were 
  reconstructed, only two Gaussian functions were used. This applies
  to 0.0--0.5~GeV$/c$ for $\Xi_c^+$, and to 4.0--4.5~GeV$/c$
  for both $\Xi_c^+$ and $\Xi_c^0$. For 4.5--5.0~GeV$/c$ no signal
  events were generated, so the fits of 4.0--4.5~GeV$/c$ were used.
} The lineshape is then fixed for fitting the data.

The fit to the data is performed in several steps:
\begin{enumerate}
\item First, we fit just the
  background function to an upper mass sideband of $\Xi_c'$, 
  $2625 < \Delta m + m_{\mathrm{offset}} < 2900$~MeV$/c^2$, using a
  binned maximum likelihood method. This provides the initial values
  of the background parameters. 
\item Next, we attempt to fit the combined background and signal
  functions in the mass range 2550--2700~MeV$/c^2$. The signal mass and
  yield are floated, with the initial value of the signal yield 
  set to zero. The background parameters are also left free; all
  other parameters are fixed.
  The mass is initially set to a central value extracted from a
  fit to the entire dataset and is allowed to vary within
  10~MeV/$c^2$ around this value. The fit uses a binned maximum
  likelihood method, followed by a binned $\chi^2$ minimization
  with MINOS error-handling enabled~\cite{minuit}.
\item We then check whether the fit converges to a physical value.
  If the signal mass is within 2~MeV$/c^2$ of the edge of the allowed range
  or the yield is unphysical (above~50,000 or below~$-100$), we
  reject the fit. This typically occurs when studying a region
  of phase space where the signal yield is too small to fit with
  a floating mass.
\item If the first fit is rejected, we reset the parameters to
  the initial values described in step 2 and fix the mass to the
  central value extracted from the fit to the entire dataset.
  The fit is then repeated.
\end{enumerate}
The individual fitted spectra are shown in Appendix~\ref{app:spectra}.

To remove the category 4 background described in 
Section~\ref{sec:background}, we perform a sideband subtraction in
$m(\Xi^- \pi^+ [\pi^+])$ as follows:
\begin{itemize}
\item The $\Delta m$ distribution of events in the
  $\Xi_c$ mass signal region $-3\sigma < m(\Xi^- \pi^+ [\pi^+]) - m_0 < +3\sigma$
  is plotted, where $\sigma$ is the $\Xi_c$ mass resolution and $m_0$
  is the central value of the $\Xi_c$ mass peak. 
\item Similarly, the $\Delta m$ distribution for events in the
  $\Xi_c$ mass sidebands,
  $5\sigma < \left|m(\Xi^- \pi^+ [\pi^+])-m_0\right| < 8\sigma$,
  is plotted.
\item The $\Delta m$ distributions are fitted with a signal lineshape
  plus a polynomial background as described above.
  The integral of the signal function gives the combined yields of
  events from categories~1 and~4 in that $m(\Xi^- \pi^+ [\pi^+])$ range.
\item The fitted yield from the sidebands is subtracted from the fitted
  yield in the signal region.
\end{itemize}
This process suppresses the category 4 background but retains the signal
with high efficiency ($\sim 95\%$ at low $p^*$ and $\sim 90\%$ at high $p^*$).
A small fraction of category 4 events have peaking
structure in $m(\Xi^- \pi^+ [\pi^+])$ and survive the sideband subtraction;
this rate is 1\% or less of the category 1 rate in all cases so we neglect it.

\begin{figure}
  \begin{center}
    \begin{tabular}{cc}
      \epsfig{file=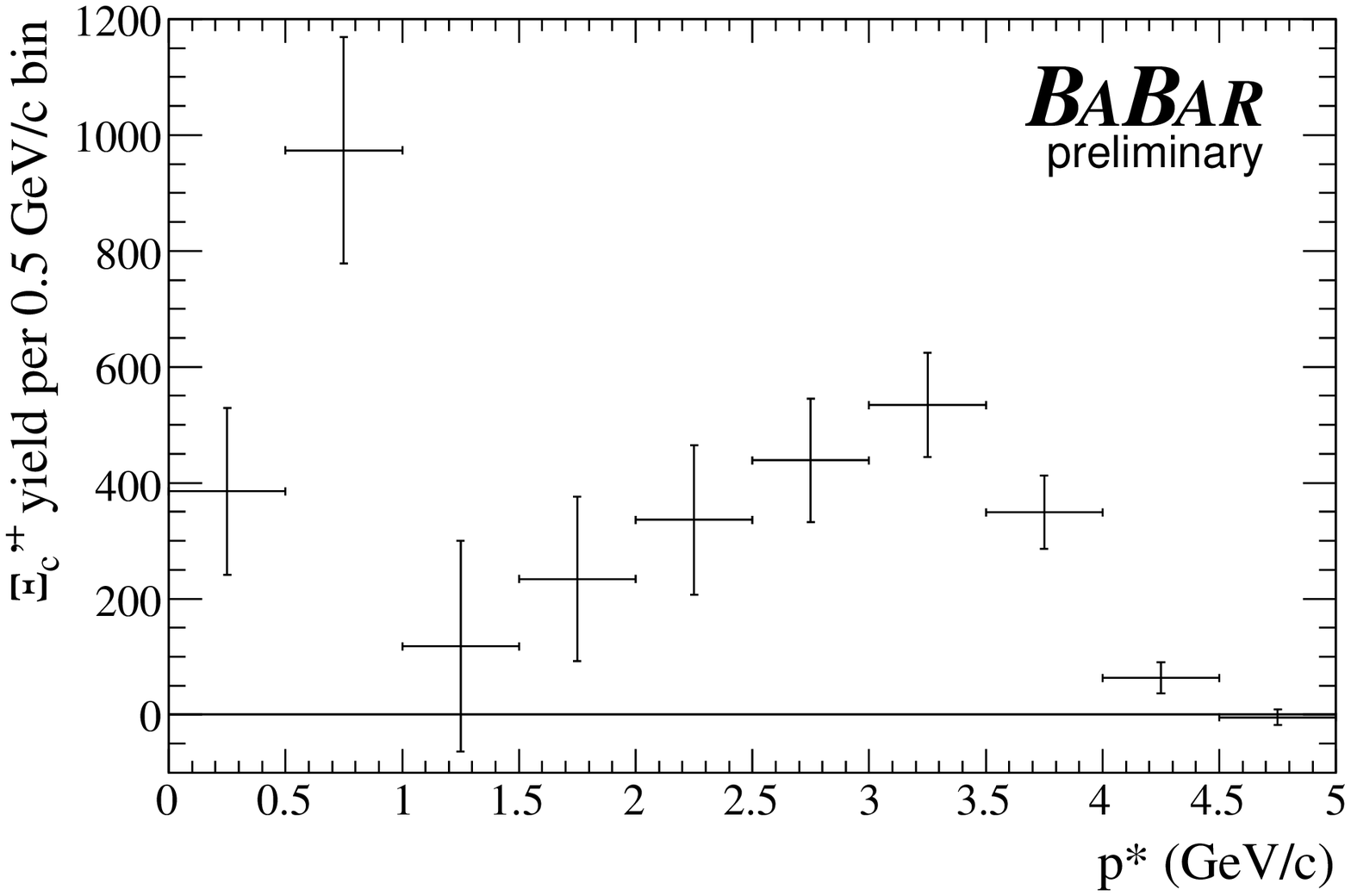, width=0.5\textwidth} &
      \epsfig{file=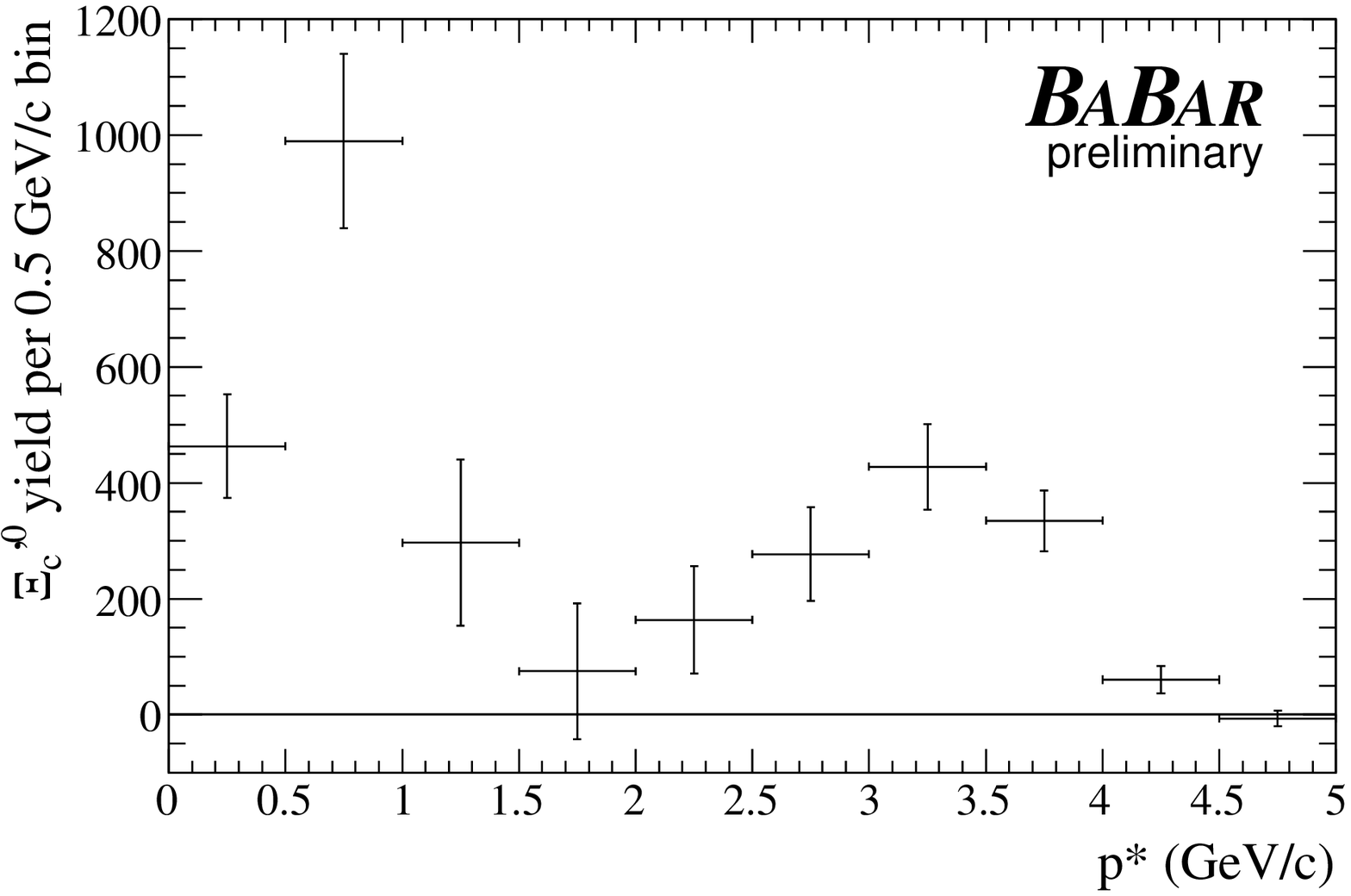, width=0.5\textwidth} 
      \begin{picture}(0.0,0.0)
	\put(-300, 105){\bf(a)}
	\put(-57, 105){\bf(b)}
      \end{picture}
    \end{tabular}
  \end{center}
  \caption[Background-subtracted $\Xi_c$ momentum spectra]
	  {Background-subtracted $\Xi_c$ momentum spectra, not corrected
	    for efficiency. The yield in each $p^*$ interval is shown 
	    for (a) $\Xi_c^{'+}$ and (b) $\Xi_c^{'0}$.
	  }
	  \label{fig:sub_pstar_spectra}
\end{figure}

After performing the sideband subtraction described above,
the numbers of background-subtracted
$\Xi_c^{'+}$ and $\Xi_c^{'0}$ are $3341 \pm 375$ and $3195 \pm 301$, respectively.
The $p^*$ distributions are shown in Fig.~\ref{fig:sub_pstar_spectra}.
As discussed in Section~\ref{sec:Introduction}, there are two
contributions: $\Xi_c'$ from $B$ decays and from the continuum.
$\Xi_c'$ produced in $B$ decays have low momentum, especially
if the recoiling antibaryon is also charmed. Allowing for the
motion of the $B$ mesons, the kinematic limit is $p^* < 2.08$~GeV$/c$,
but this corresponds to the process $\Bb \rightarrow \Xi_c' \antiproton$
which is heavily suppressed. The limit for Cabibbo-allowed
processes is $p^* < 2.02$~GeV$/c$. By contrast, continuum production
occurs mainly at higher values of $p^*$, with a kinematic limit of
$p^* < 4.63$~GeV$/c$ at $\sqrt s = 10.6$~GeV. Two separate peaks corresponding
to these processes are clearly visible in 
Fig.~\ref{fig:sub_pstar_spectra}.

\subsection{Efficiency correction}

\begin{figure}
\begin{center}
\begin{tabular}{cc}
\epsfig{file=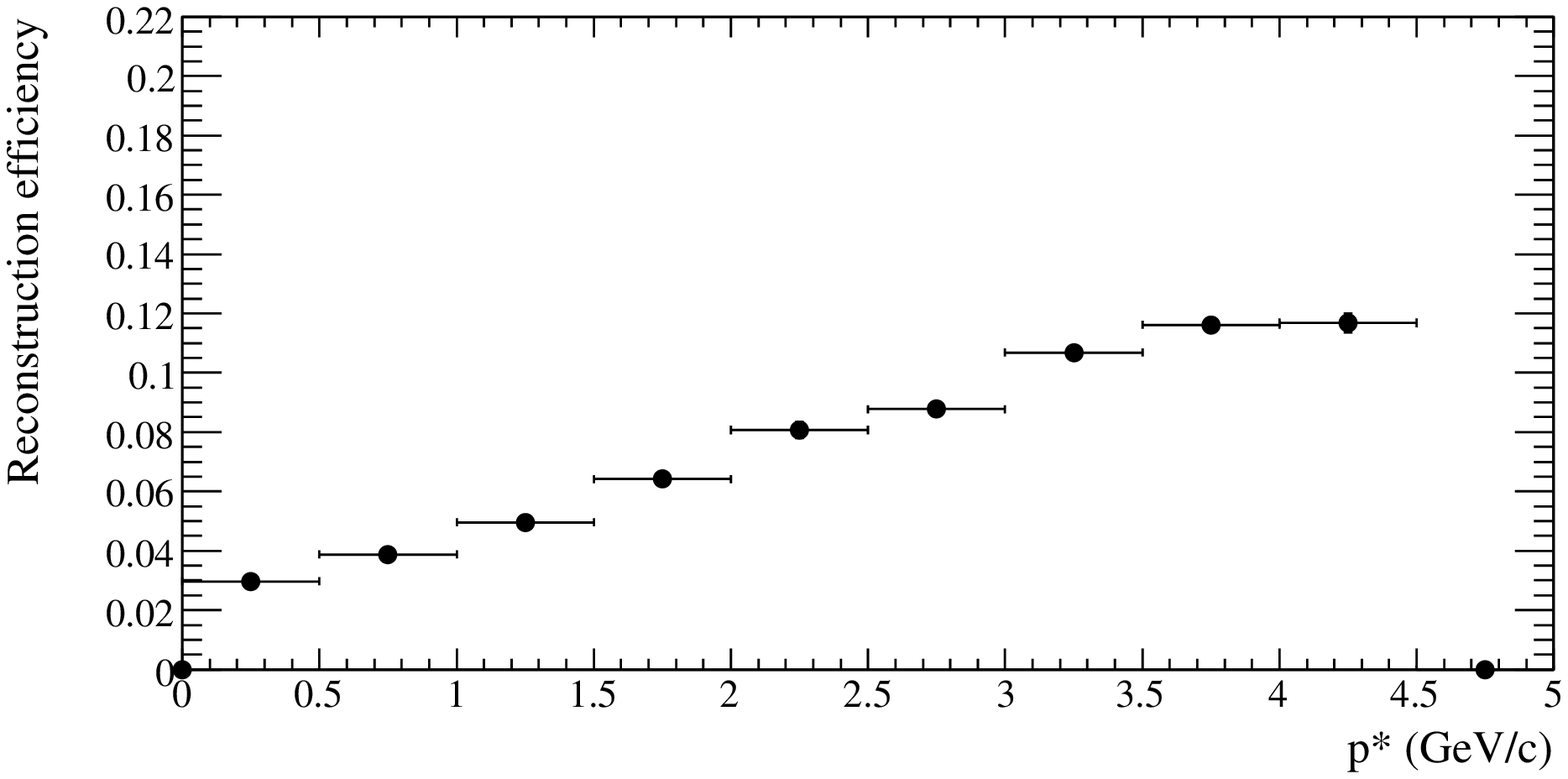,width=0.5\textwidth} &
\epsfig{file=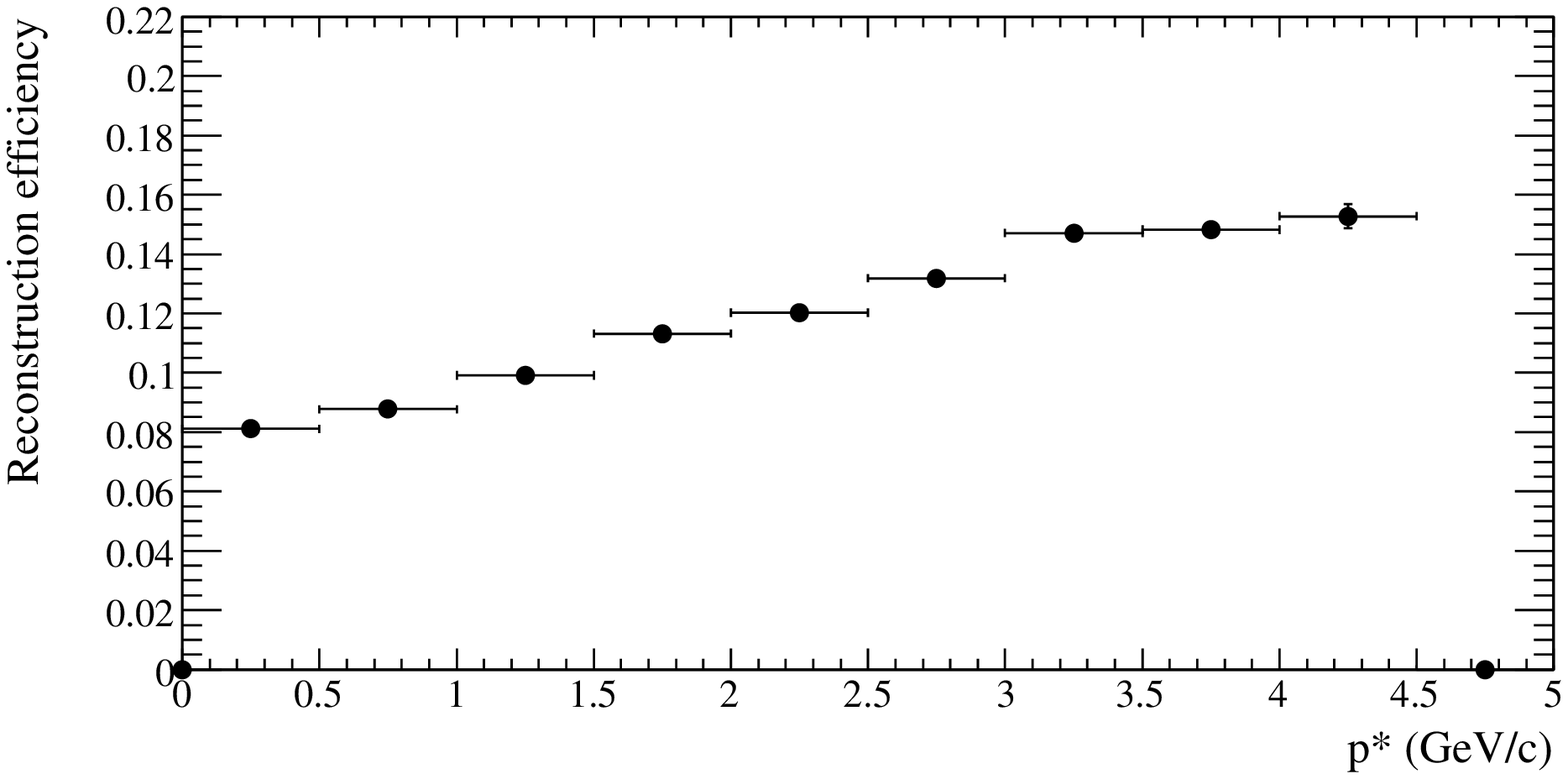,width=0.5\textwidth}
      \begin{picture}(0.0,0.0)
        \put(-300, 95){\bf(a)}
        \put(-57, 95){\bf(b)}
      \end{picture}
    \end{tabular}
  \end{center}
  \caption[Efficiency]
          {Efficiency as a function of $p^*$ for (a) $\Xi_c^{'+}$ and
           (b) $\Xi_c^{'0}$. The factor $\mathcal{B}(\Lambda \rightarrow p \pi^-)$
           is not included.
          }
  \label{fig:efficiency}
\end{figure}

For each $p^*$ interval,
the efficiency $\varepsilon$ is determined from simulated
events in the corresponding $p^*$ range: 
\begin{equation}
  \label{eq:effic}
  \varepsilon = 
  \frac{
    \mbox{Yield of true $\Xi_c'$ in $m(\Xi^- \pi^+ [\pi^+])$ signal window}
  }{
    \mbox{Number of generated $\Xi_c'$}
  }
  ,
\end{equation}
where the yield is obtained by fitting the $\Delta m$ spectrum
and true $\Xi_c'$ are identified with MC generator information.
An additional correction is made to take into account signal events
which fall into the $m(\Xi^- \pi^+ [\pi^+])$ sidebands and are subtracted from
the yield as described in Section~\ref{sec:fit}. This
correction, $\delta \varepsilon$, 
is obtained from the $m(\Xi^- \pi^+ [\pi^+])$ lineshape of simulated $\Xi_c'$
in the relevant $p^*$ range:
\begin{equation}
  \label{eq:efficcorrection}
    \delta \varepsilon =
    \frac{
      \mbox{Yield of true $\Xi_c'$ in $m(\Xi^- \pi^+ [\pi^+])$ sideband}
    }{
      \mbox{Yield of true $\Xi_c'$ in $m(\Xi^- \pi^+ [\pi^+])$ signal window}
   }
   .
\end{equation}
This is a small effect (approximately 1\% for $\Xi_c'$ produced in $B$
decays and 2\% for $\Xi_c'$ produced from the continuum).
The overall efficiency, $\varepsilon (1 - \delta \varepsilon)$,
is shown in Fig.~\ref{fig:efficiency}.

\section{SYSTEMATIC STUDIES}
\label{sec:Systematics}

The following systematic effects are considered. The first four
are applied only to the overall
normalization; the others are treated as fully uncorrelated
and are applied to each $p^*$ interval separately.
All uncertainties quoted are relative.
\begin{description}
  \item[Particle identification:] A systematic uncertainty of 3.5\%
    is assigned to the efficiency of the proton identification,
    as in a previous analysis of the $\Xi_c$ system~\cite{babar_xic}.
  \item[Tracking efficiency:] To correct for a known discrepancy
    in tracking efficiency between data and simulation, systematic corrections of
    $(2.35 \pm 7.0)\%$ and $(1.55 \pm 5.6)\%$ are applied to the
    efficiency for reconstruction of  
    $\Xi_c^+ \rightarrow \Xi^- \pi^+ \pi^+$ and
    $\Xi_c^0 \rightarrow \Xi^- \pi^+$, respectively.
  \item[Photon efficiency:] Based on studies of photon-finding
    efficiency in control samples, a systematic uncertainty of 1.8\% is
    applied to the efficiency.
  \item[$\boldmath{\Lambda}$ branching fraction:] The world-average branching
    fraction is $\mathcal{B} (\Lambda \rightarrow p \pi^-)=(63.9 \pm 0.5)\%$.
    This results in a 0.8\% systematic uncertainty.
  \item[Finite simulation statistics:] The statistical uncertainty
    in the efficiency calculation in each $p^*$ interval is applied 
    as a systematic uncertainty to the specific data point. This is
    5\% or lower in each interval.
  \item[Signal fitting procedure:] The analysis is repeated with a
    different functional form for the signal lineshape, described in
    Section~\ref{appendix:lineshape:nov}
    of Appendix~\ref{appendix:lineshape}.
    For each $p^*$ interval, the systematic uncertainty is taken to be
    the difference between the efficiency-corrected
    yields from the two functional forms divided by $\sqrt 2$.
    The value depends on the specific $p^*$ interval, but is typically
    around 5--10\%.
  \item[Background fitting procedure:] The background shape is changed
    from a second-order polynomial to a fourth-order polynomial and
    the fit range is increased substantially. For each
    $p^*$ interval, the systematic uncertainty is taken to be
    the difference between the efficiency-corrected
    yields from the two methods divided by $\sqrt 2$.
    This varies between $p^*$ intervals, but is typically
    around 5--10\%. In a few intervals with low yields it
    rises to 30--50\%, but is always lower than the statistical
    uncertainty.
  \item[Efficiency correction within a $p^*$ interval:] If the simulation
    does not correctly model the $p^*$ distribution within a $p^*$
    interval and the efficiency varies significantly across that
    interval, the efficiency may not be predicted correctly. 
    This effect was studied in a previous analysis and found to
    be a few percent or less with the \babar\ simulation~\cite{babar_xic}.
    We assign a systematic uncertainty of 4\% for each $p^*$ interval.
  \item[Intermediate resonances in $\Xi_c^+ \rightarrow \Xi^- \pi^+ \pi^+$:]
    In the simulation, the $\Xi_c^+$ three-body decay is assumed
    to be entirely non-resonant. However, structure is observed in
    the Dalitz plot distributions in data. We determine the efficiency for 
    two extreme cases:
    when the non-resonant contribution is 0\% ($\varepsilon_{\mathrm{res}}$)
    and when the non-resonant contribution is 100\%
    ($\varepsilon_{\mathrm{nonres}}$). The overall efficiency 
    is then changed to 
    $(\varepsilon_{\mathrm{res}} + \varepsilon_{\mathrm{nonres}}) / 2$
    with a systematic uncertainty of
    $(|\varepsilon_{\mathrm{res}} - \varepsilon_{\mathrm{nonres}}|) / 2$.
    This only affects the $\Xi_c^{'+}$ mode. The effect is approximately
    15\% at low $p^*$, dropping to zero at high $p^*$ as the efficiency
    becomes more uniform across the Dalitz plot.
  \item[Finite resolution:] The reconstructed $p^*$ distribution is
    the true $p^*$ distribution convoluted by the resolution
    function. The resolution varies with $p^*$, but is
    typically 15--20~MeV$/c$, substantially smaller than
    the bin size (500~MeV$/c$). We therefore neglect this effect.
  \item[Admixture of \ccbar and \BB simulation:] The angular distributions
    of $\Xi_c'$ produced in continuum events and in $B$ decays differ,
    resulting in slightly different efficiencies for the two processes.
    This is modelled in the simulation. In general it is unambiguous
    which process dominates in a given $p^*$ interval, but in the $p^*$
    interval 1.5--2.0~GeV$/c$ both may contribute significantly, leading
    to a slight dependence of the efficiency-corrected yield on the
    assumed relative production rates. However, the absolute yield in this
    interval is small and other uncertainties dominate, so we neglect this
    effect.
\end{description}

\begin{figure}
  \begin{center}
    \begin{tabular}{cc}
\epsfig{file=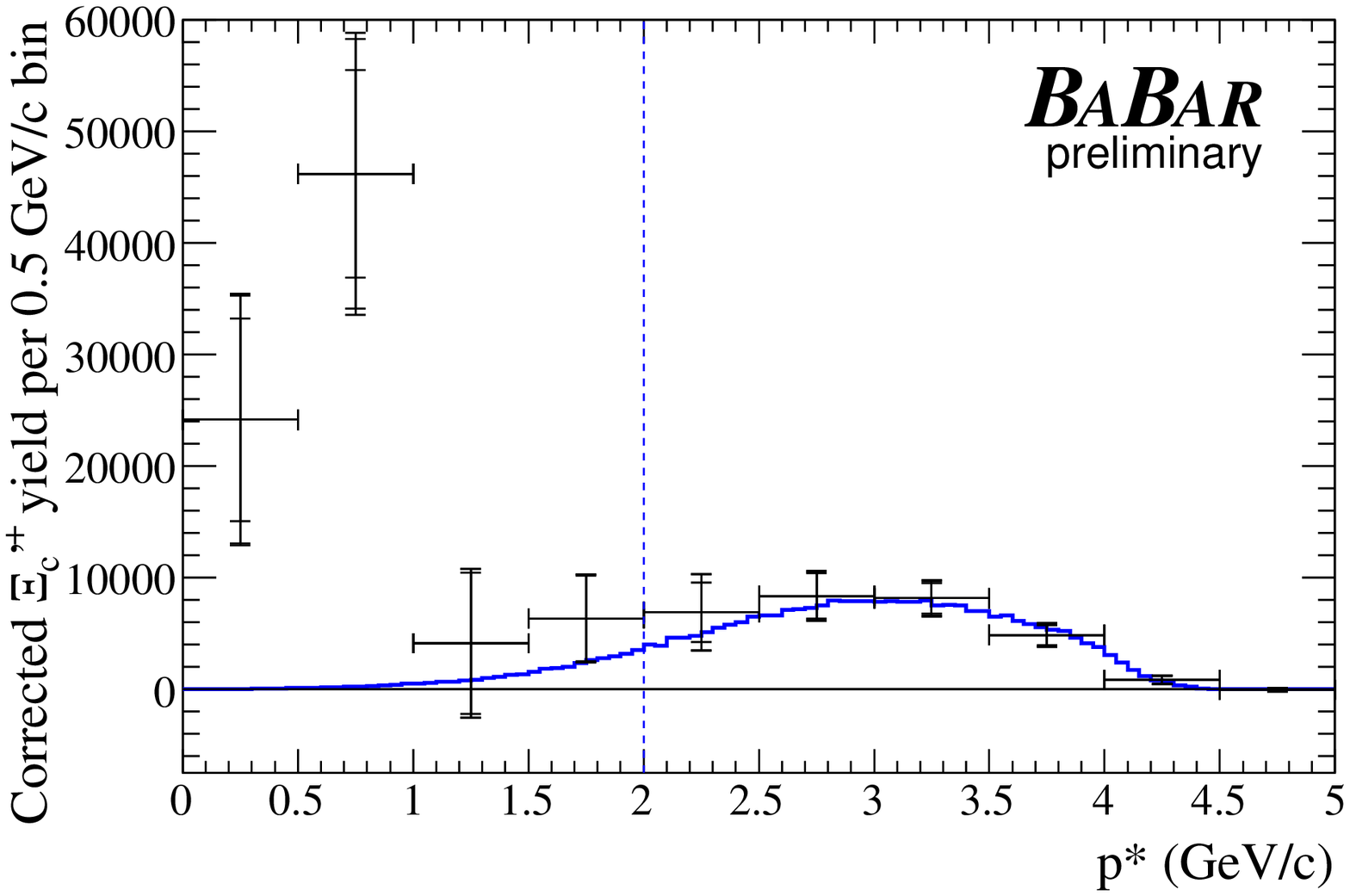, width=0.5\textwidth} &
\epsfig{file=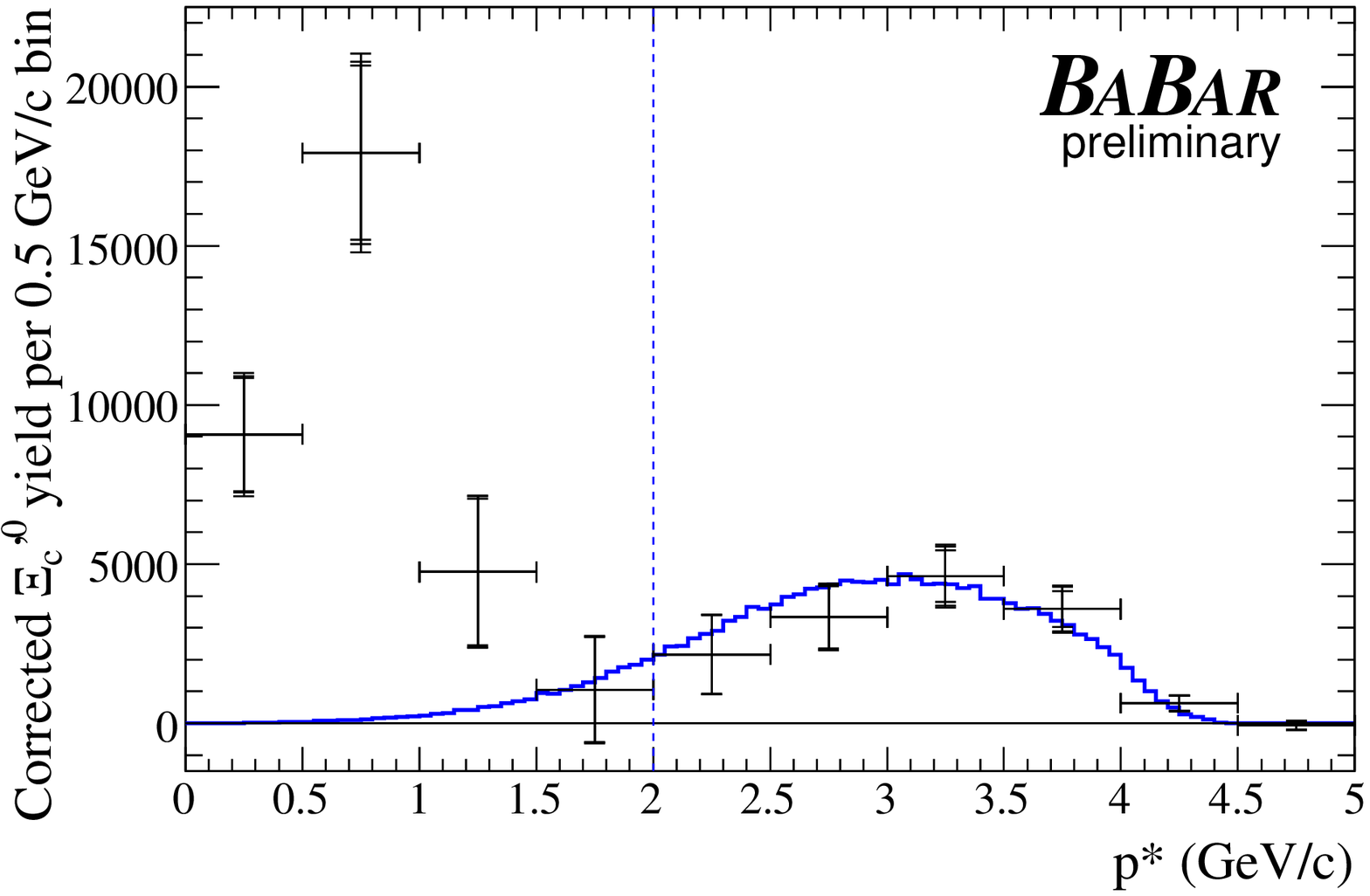, width=0.5\textwidth} 
      \begin{picture}(0.0,0.0)
	\put(-300, 105){\bf(a)}
	\put(-57, 105){\bf(b)}
      \end{picture}
    \end{tabular}
  \end{center}
  \caption[Efficiency-corrected, background-subtracted $p^*$ spectrum]
	  {The efficiency-corrected, background-subtracted $p^*$ spectrum
	    for (a) $\Xi_c^{'+}$ and (b) $\Xi_c^{'0}$. The curve is the
	    simulated continuum distribution described in 
	    Section~\ref{sec:Physics}; it is fitted to the data for
	    $2.0 < p^* < 4.5$~GeV$/c$ (indicated by the dashed line).
	  }
	    \label{fig:fully_corrected_spectrum}
\end{figure}

After applying these systematic corrections and uncertainties, the
$p^*$ spectrum shown in Fig.~\ref{fig:fully_corrected_spectrum} is obtained.
The inner error bars show the statistical uncertainty
(both from data and simulation), the middle error bars show the
sum in quadrature of the statistical and uncorrelated systematic
uncertainties, and the outer error bars (where visible) show the
sum of all uncertainties in quadrature.

\section{PHYSICS RESULTS}
\label{sec:Physics}

\subsection{Production rates}

It is clear from Fig.~\ref{fig:fully_corrected_spectrum} that there is
significant $\Xi_c'$ production both in $B$ decays and from the \ccbar
continuum.
We separate the contributions of the two processes as follows:
\begin{itemize}
  \item $B$ production of $\Xi_c'$ for $p^* > 2.0$~GeV$/c$ is assumed to be zero.
  \item Continuum production of $\Xi_c'$ for $p^* > 2.0$~GeV$/c$ is taken
    to be the sum of the measured yields in each $p^*$ interval above 2.0~GeV$/c$.
  \item The data between 2.0~GeV$/c$ and 4.5~GeV$/c$ are then fitted with a suitable 
    function, described below. The function is extrapolated down to $p^* = 0$ and the
    integral over the range 0.0--2.0~GeV$/c$ is taken as the
    continuum production of $\Xi_c'$ in that momentum range.
  \item The $B$ production of $\Xi_c'$ below 2.0~GeV$/c$ is taken to
    be the sum of the measured yields in each $p^*$ interval below 2.0~GeV$/c$
    less the estimated continuum production in the range
    0.0--2.0~GeV$/c$
\end{itemize}
The continuum function is based on the Bowler fragmentation model~\cite{ref:bowler},
tuned to the global \babar\ data and implemented within the
JETSET~\cite{ref:pythia}
generator. Only the amplitude is allowed to float in the fit to the
data. The fitted function is shown in Fig.~\ref{fig:fully_corrected_spectrum}.
The $\chi^2/\mathrm{NDF}$ of the fits are $1.2/4$ and $1.8/4$
for $\Xi_c^{'+}$ and $\Xi_c^{'0}$, respectively.

\begin{figure}
\begin{center}
    \begin{tabular}{cc}
      \epsfig{file=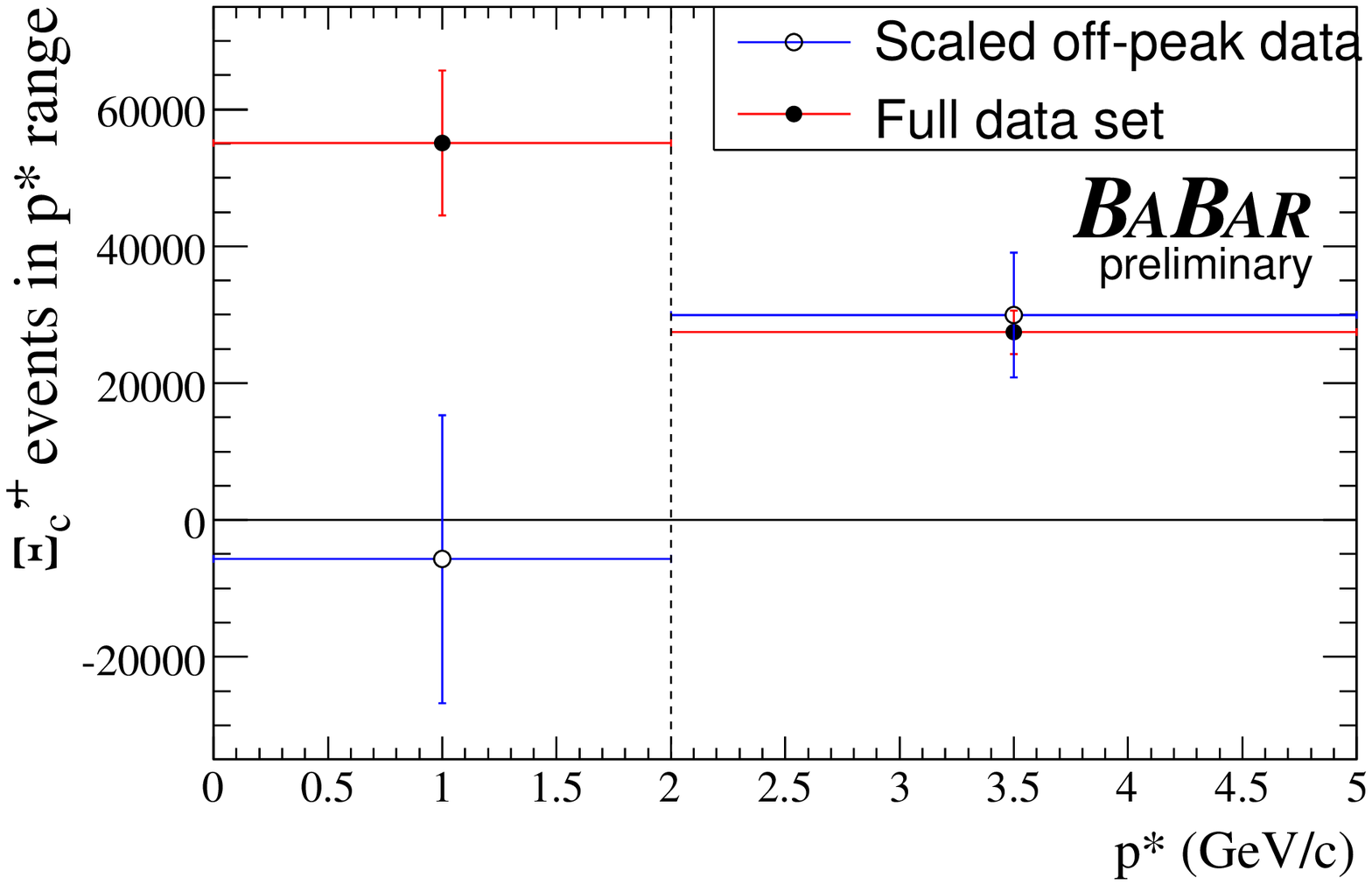, width=0.48\textwidth} &
      \epsfig{file=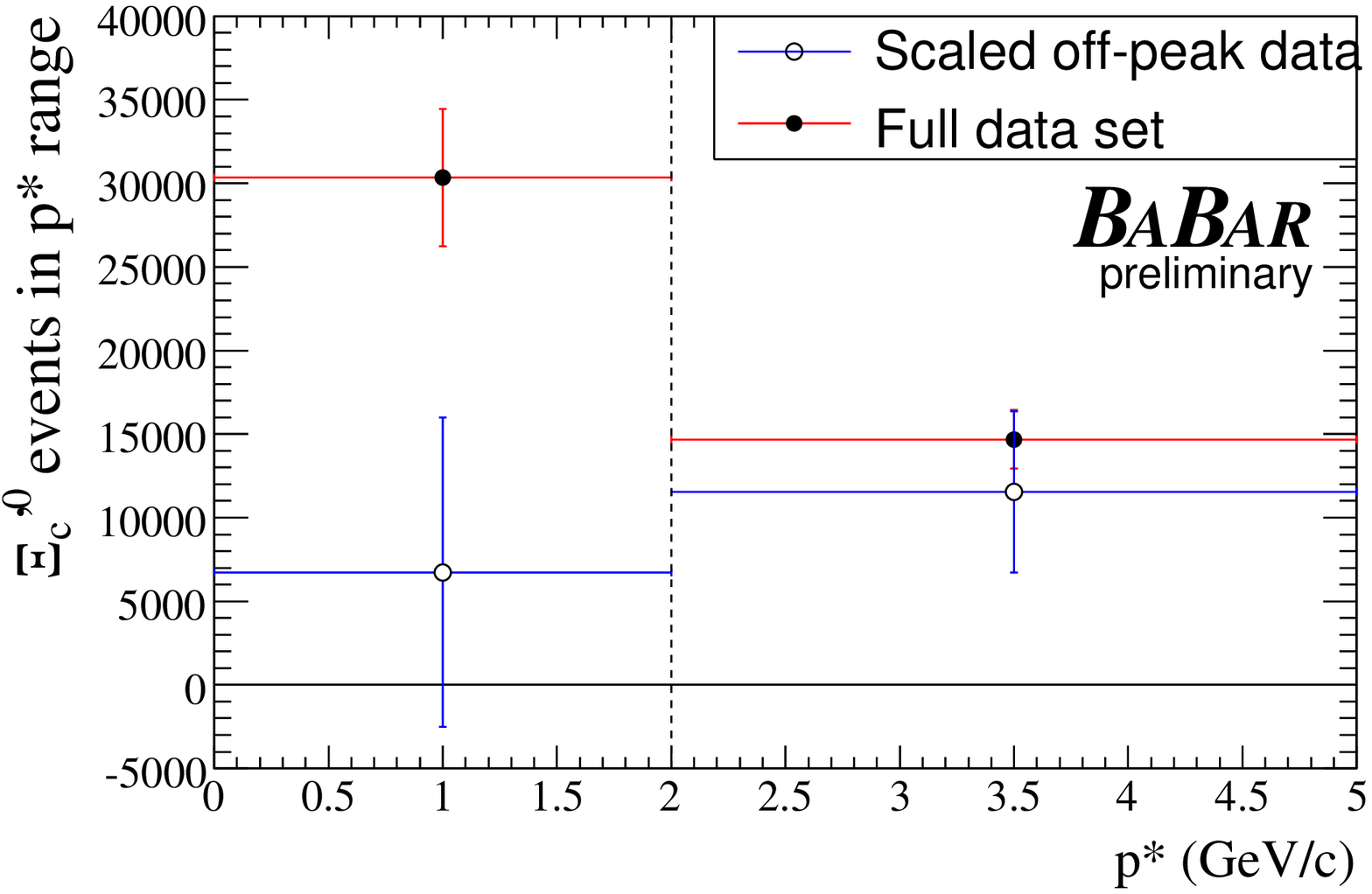, width=0.48\textwidth} 
      \begin{picture}(0.0,0.0)
	\put(-290, 35){\bf(a)}
	\put(-45, 48){\bf(b)}
      \end{picture}
    \end{tabular}
  \end{center}
  \caption[Comparison of yields in the off-peak sample with the full sample]
          {
            Comparison of the efficiency corrected yields in the off-peak
            sample with those of the full data set (off-peak and on-peak
            combined). The off-peak yields have been scaled up to account
            for the difference in integrated luminosity, and corrected for
            the small change in the continuum cross-section with $\sqrt s$.
            Plot~(a) shows the $\Xi_c^{'+}$ and
            plot~(b) shows the $\Xi_c^{'0}$.
            Systematic uncertainties and corrections are not included.
          }
          \label{fig:offpeak}
\end{figure}

To test the model dependency, we fit the data with a
number of other fragmentation models.
For these crosschecks, we take the parameterizations
to be functions of the scaled momentum
$x_p$. The data above 2.0~GeV$/c$ are
well-described by the Peterson model~\cite{ref:peterson}
and by a baryon-specific version of the phenomenological model of
Kartvelishvili et al.~\cite{ref:kpmb}.
The standard deviation of the extracted rates when using these
three fragmentation models is quoted as the model-dependent uncertainty:
\begin{center}
  \begin{tabular}{lc}
    $\Xi_c^{'+}$ from continuum: & 32681 $\pm$ 5516 (exp.) $\pm$ 4443 (model)\\
    \noalign{\vskip1pt}
    $\Xi_c^{'+}$ from $B$ decay: & 77199 $\pm$ 7907 (exp.) $\pm$ 4443 (model)\\
    \noalign{\vskip1pt}
    $\Xi_c^{'0}$ from continuum: & 16356 $\pm$ 2509 (exp.) $\pm$ 1384 (model)\\
    \noalign{\vskip1pt}
    $\Xi_c^{'0}$ from $B$ decay: & 30782 $\pm$ 3088 (exp.) $\pm$ 1384 (model),
  \end{tabular}
\end{center}
where the experimental uncertainties combine both statistical and systematic
effects.
Excluding the normalization systematic uncertainties, these correspond to
a statistical significance for $\Xi_c'$ production in $B$ decays in excess
of $12\sigma$ for each mode, and a significance for continuum production
at $p^*>2.0$~GeV$/c$ in excess of $6\sigma$ for each mode.
As an additional crosscheck, the continuum production for
$p^*<2.0$~GeV$/c$ is measured in the off-peak data sample alone.
This procedure is model-independent (to 2--3\%) but has a much
larger statistical uncertainty. The results are shown in
Fig.~\ref{fig:offpeak}
and are consistent with
the yields quoted above within statistical uncertainties.

Dividing the above yields by twice the total number of $\BB$ pairs in the data 
sample, we measure the product branching fractions as:
\begin{eqnarray*}
  \mathcal{B}(B \rightarrow \Xi_c^{'+} X) \times
  \mathcal{B}(\Xi_c^+ \rightarrow \Xi^- \pi^+ \pi^+)
  &=& [ 1.69 \pm 0.17~\mathrm{(exp.)} \pm 0.10~\mathrm{(model)} ] \times 10^{-4} \\
  \mathcal{B}(B \rightarrow \Xi_c^{'0} X) \times
  \mathcal{B}(\Xi_c^0 \rightarrow \Xi^- \pi^+)
  &=& [ 0.67 \pm 0.07~\mathrm{(exp.)} \pm 0.03~\mathrm{(model)} ] \times 10^{-4}
  .
\end{eqnarray*}
Comparing the second measurement with a previous \babar\ result~\cite{babar_xic},
\begin{displaymath}
  \mathcal{B}(B \rightarrow \Xi_c^{0} X) \times
  \mathcal{B}(\Xi_c^0 \rightarrow \Xi^- \pi^+)
  = [ 2.11 \pm 0.19 ~\mathrm{(stat.)} \pm 0.25~\mathrm{(sys.)} ] \times 10^{-4},
\end{displaymath}
we observe that approximately one third of $\Xi_c^0$ produced in $B$ decays
come from $\Xi_c^{'0}$ decays.

Correcting for the $1/s$ scaling of the continuum cross-section for
data taken at $\sqrt s = 10.54$~GeV
and taking into account the 1.5\% systematic uncertainty on the integrated
luminosities quoted in Section~\ref{sec:babar},
the cross-sections at $\sqrt s = 10.58$~GeV are:
\begin{eqnarray*}
  \sigma(e^+ e^- \rightarrow \Xi_c^{'+} X) \times \mathcal{B}(\Xi_c^+ \rightarrow \Xi^- \pi^+ \pi^+)
  &=& 141 \pm 24 \ \mathrm{(exp.)} \pm 19 \ \mathrm{(model)} ~\mathrm{fb} \\
  \sigma(e^+ e^- \rightarrow \Xi_c^{'0} X) \times \mathcal{B}(\Xi_c^0 \rightarrow \Xi^- \pi^+)
  &=& 70 \pm 11 \ \mathrm{(exp.)} \pm 6 \ \mathrm{(model)} ~\mathrm{fb}
  .
\end{eqnarray*}
Comparing this with a previous \babar\ result~\cite{babar_xic}:
\begin{displaymath}
  \sigma(e^+ e^- \rightarrow \Xi_c^0 X) \times \mathcal{B}(\Xi_c^0 \rightarrow \Xi^- \pi^+)
  = 388 \pm 39 ~\mathrm{(stat.)} \pm 41~\mathrm{(sys.)} ~\mathrm{fb}
\end{displaymath}  
we observe that about 18\% of $\Xi_c^0$ produced in the
continuum come from a $\Xi_c^{'0}$.

\subsection{Helicity angle distribution}
\label{sec:helicity}

The quark-model predicts $J^P = \frac{1}{2}^+$
for $\Xi_c'$, $\Xi_c$ and $\Xi^-$. Under this assumption, the
helicity angle distribution for the decay processes studied
should be flat in the cosine of the helicity angle, $\cos\theta_h$,
where $\theta_h$ defined as is the angle between the $\Xi^-$ direction in the
$\Xi_c$ rest frame and the $\Xi_c$ direction in the $\Xi_c'$ rest frame.
If the $\Xi_c'$ has $J = \frac{3}{2}$, the angular distribution is of the
form $c_1 + c_2 \cos^2\theta_h$ where $c_1$ and $c_2$ are unknown
parameters and $c_2$ may be zero. 
In general the distribution for higher spins is a higher-order
polynomial---but the distribution may be flat if the
higher-order coefficients are zero. Therefore, a non-flat
distribution in the data would exclude $J = \frac{1}{2}$,
but a flat distribution would not exclude $J > \frac{1}{2}$.

\begin{figure}
  \begin{center}
      \epsfig{file=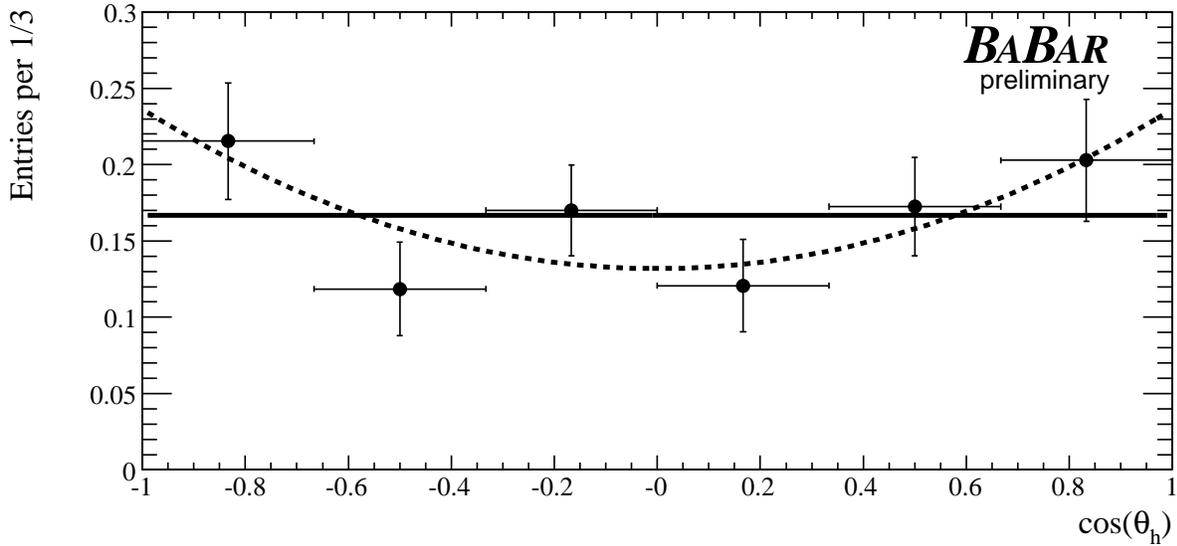, width=1.0\textwidth}
  \end{center}
  \caption[Fits to helicity distributions]
          {Fits to efficiency-corrected helicity distributions
            for data with $p^* > 2.5$~GeV$/c$.
            The plot shows the normalized, weighted sum of the $\Xi_c^{'+}$
	    and~$\Xi_c^{'0}$ distributions. The solid line
	    assumes a flat helicity distribution, whereas for the
	    dashed line a quadratic term is added.

          }
  \label{fig:helicity:fits}
\end{figure}

The method used to measure the helicity angle distribution in
the data is similar to the one used for the $p^*$
spectrum: we divide the data into six slices of $\cos \theta_h$
and fit the mass spectrum in each slice. A $p^*$ threshold of
2.5~GeV$/c$ is applied throughout to improve the
signal-to-background ratio. After correcting for efficiency,
the distributions are fitted with two functions: first,
a flat distribution:
\begin{equation}
  \label{eq:flat}
  f_0 (\cos\theta_h) = \alpha
\end{equation}
and second, a symmetric quadratic:
\begin{equation}
  \label{eq:quad}
  f_0 (\cos\theta_h) = \alpha \left( 1 + \beta \cos^2\theta_h \right)
  .
\end{equation}
Separate fits to each mode in the data are performed
and the fit results are shown in
Table~\ref{tab:helicity:fits}.
The data are clearly consistent with being flat
($\chi^2/\mathrm{NDF}$ less than unity). The fitted quadratic
parameters $\beta$ are consistent with zero, though with
large statistical uncertainties. Since the two modes
should have identical helicity distributions, we weight
them according to their statistical precision and combine
them---this is shown in
Fig.~\ref{fig:helicity:fits}
and the fit results are given in Table~\ref{tab:helicity:fits}.
The data are still fit well by a flat distribution, though
with a somewhat reduced $\chi^2$ probability (20\%).
From this we conclude that the data are consistent with the
predicted $J=\frac{1}{2}$ but that higher spins cannot be excluded.

\begin{table}
  \caption[Results of fits to helicity distributions]
          {Results of fits to efficiency-corrected helicity distributions.
	    The $\chi^2$ goodness-of-fit and the number of degrees of
	    freedom are given for the flat and quadratic distributions
	    given in Eq.~\ref{eq:flat} and~\ref{eq:quad}, along with
	    the fitted parameter $\beta$ from the quadratic distribution.
	    The results are given for the $\Xi_c^{'+}$ and $\Xi_c^{'0}$
	    samples individually, and for a weighted sum of the two samples.
	  }
  \begin{center}
    \begin{tabular}{lccc}
                                 &   $\Xi_c^{'+}$   &   $\Xi_c^{'0}$   & Weighted Sum\\
      \hline \noalign{\vskip1pt}
      $\chi^2/\mathrm{NDF}$ for flat      & $4.0/5$ $(55\%)$ & $4.4/5$ $(50\%)$ & $7.3/5$ $(20\%)$ \\   \noalign{\vskip1pt}
      $\chi^2/\mathrm{NDF}$ for quadratic & $2.7/4$ $(61\%)$ & $1.3/4$ $(87\%)$ & $3.0/4$ $(55\%)$ \\   \noalign{\vskip1pt}
      $\beta$ for quadratic      & $0.63 \pm 0.68$  & $1.04 \pm 0.82$  & $0.79 \pm 0.52$    \\   \noalign{\vskip1pt} \hline
    \end{tabular}
  \end{center}
  \label{tab:helicity:fits}
\end{table}

\section{SUMMARY}
\label{sec:Summary}

We have confirmed the CLEO observation of the $\Xi_c^{'+}$ and $\Xi_c^{'0}$
states from the \ccbar continuum. In addition, we found that
$B$ mesons decay at a substantial rate to $\Xi_c^{'+}$ and $\Xi_c^{'0}$.
This is the first observation of such decays; the statistical significance
is in excess of $12\sigma$ for each mode.
We have measured the production rates of $\Xi_c'$ from $B$
decays (expressed as a branching fraction) and from the
\ccbar continuum (expressed as a cross-section); 
in both cases, the absolute rate is scaled by the
unknown absolute $\Xi_c$ branching fraction.
We have measured the angular distribution of 
$\Xi_c' \rightarrow \Xi_c \gamma$ decays and found it to be
consistent with the prediction for $J^P = \frac{1}{2}^+$.
However, higher spins cannot be ruled out.

\section{ACKNOWLEDGMENTS}
\label{sec:Acknowledgments}

We are grateful for the 
extraordinary contributions of our \pep2\ colleagues in
achieving the excellent luminosity and machine conditions
that have made this work possible.
The success of this project also relies critically on the 
expertise and dedication of the computing organizations that 
support \babar.
The collaborating institutions wish to thank 
SLAC for its support and the kind hospitality extended to them. 
This work is supported by the
US Department of Energy
and National Science Foundation, the
Natural Sciences and Engineering Research Council (Canada),
Institute of High Energy Physics (China), the
Commissariat \`a l'Energie Atomique and
Institut National de Physique Nucl\'eaire et de Physique des Particules
(France), the
Bundesministerium f\"ur Bildung und Forschung and
Deutsche Forschungsgemeinschaft
(Germany), the
Istituto Nazionale di Fisica Nucleare (Italy),
the Foundation for Fundamental Research on Matter (The Netherlands),
the Research Council of Norway, the
Ministry of Science and Technology of the Russian Federation, 
Ministerio de Educaci\'on y Ciencia (Spain), and the
Particle Physics and Astronomy Research Council (United Kingdom). 
Individuals have received support from 
the Marie-Curie IEF program (European Union) and
the A. P. Sloan Foundation.

\newpage

\appendix

\begin{center}
  {\Large \bf Appendix}
\end{center}

\section{Lineshape parameterizations}
\label{appendix:lineshape}

\subsection{Triple Gaussian function}
\label{appendix:lineshape:gaussian}

The lineshape function $f(\Delta m)$ is parameterized as follows:
\begin{equation}
  f(\Delta m) 
  = N (1-f_2-f_3) G(\Delta m; \mu_1, \sigma_1)
  + N f_2 G(\Delta m; \mu_1 + \Delta \mu_2, \sigma_2)
  + N f_3 G(\Delta m; \mu_1 + \Delta \mu_3, \sigma_3)
\end{equation}
where $G(x; \mu, \sigma)$ is a Gaussian function of unit area
with mean $\mu$ and width $\sigma$. The nine parameters are
then interpreted as follows:
\begin{center}
  \begin{tabular}{cl}
    $N$     & Total fitted yield \\
    $f_2$   & Fraction of yield in second Gaussian function \\
    $f_3$   & Fraction of yield in third Gaussian function \\
    $\mu_1$ & Signal mass \\ 
    $\Delta \mu_2$ & Mean of second Gaussian function with respect to the signal mass\\
    $\Delta \mu_3$ & Mean of third Gaussian function with respect to the signal mass\\
    $\sigma_1$ & Width of first Gaussian function \\
    $\sigma_2$ & Width of second Gaussian function \\
    $\sigma_3$ & Width of third Gaussian function 
  \end{tabular}
\end{center}
When fitting simulated events to determine the lineshape, all nine
parameters are allowed to vary independently. In order to improve
fit convergence, the following bounds are placed on the variation
of the parameters:
\begin{center}
  \begin{tabular}{cl}
    $N$ & No bounds \\
    $f_2$ & No bounds \\
    $f_3$ & No bounds \\
    $\mu_1$ & Limited to $(-5,+5)$~MeV$/c^2$ relative to the true mass \\
    $\Delta \mu_2$ & Limited to $(-15,+5)$~MeV$/c^2$ relative to $\mu_1$ \\
    $\Delta \mu_3$ & Limited to $(-35,-5)$~MeV$/c^2$ relative to $\mu_1$ \\
    $\sigma_1$ & Limited to $(0,10)$~MeV$/c^2$ \\
    $\sigma_2$ & Limited to $(4,20)$~MeV$/c^2$ \\
    $\sigma_3$ & Limited to $(20,100)$~MeV$/c^2$ 
  \end{tabular}
\end{center}

\subsection{Alternative lineshape parameterization}
\label{appendix:lineshape:nov}

The following functional form was also used as a
cross-check and to determine the systematic uncertainty
due to the signal lineshape:
\begin{center}
  \epsfig{file=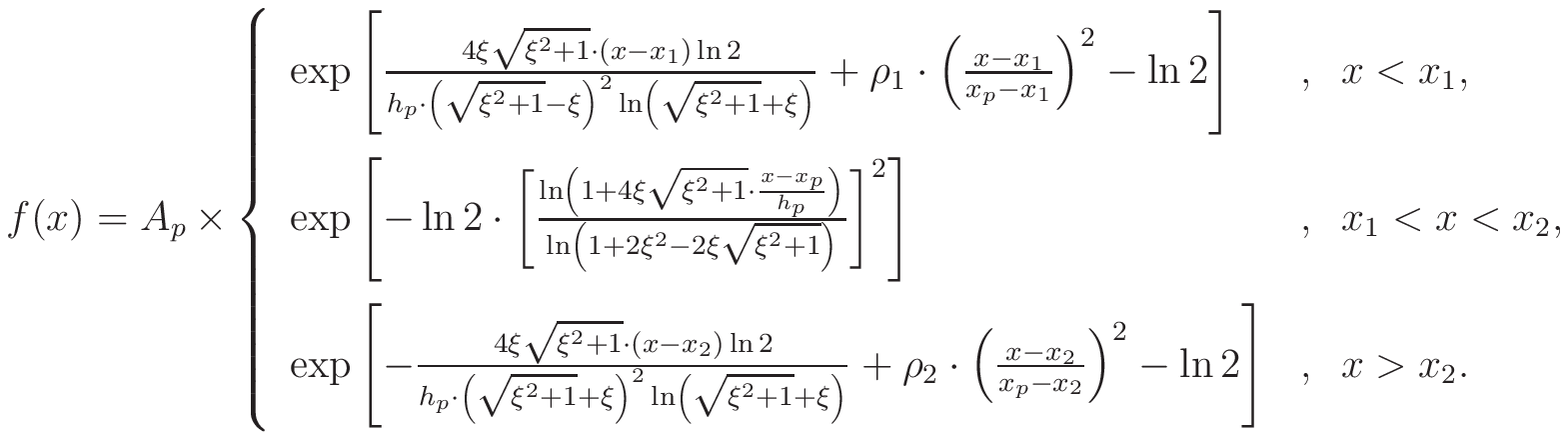, width=0.8\textwidth}
\end{center}
where
\begin{eqnarray*}
  x_1 &\equiv& x_p + \frac{h_p}{2} \left[ \frac{\xi}{\sqrt{\xi^2+1}} - 1 \right] \\
  x_2 &\equiv& x_p + \frac{h_p}{2} \left[ \frac{\xi}{\sqrt{\xi^2+1}} + 1 \right] 
  .
\end{eqnarray*}
Of the six parameters 
  $A_p$ controls the amplitude,
  $x_p$ controls the peak position,
  $h_p$ controls the width,
  $\xi$ controls the asymmetry in the central region,
  $\rho_1$ controls the lower tail, and
  $\rho_2$ controls the upper tail.

\section{Individual mass spectra}
\label{app:spectra}

\begin{figure}
  \begin{center}
    \epsfig{file=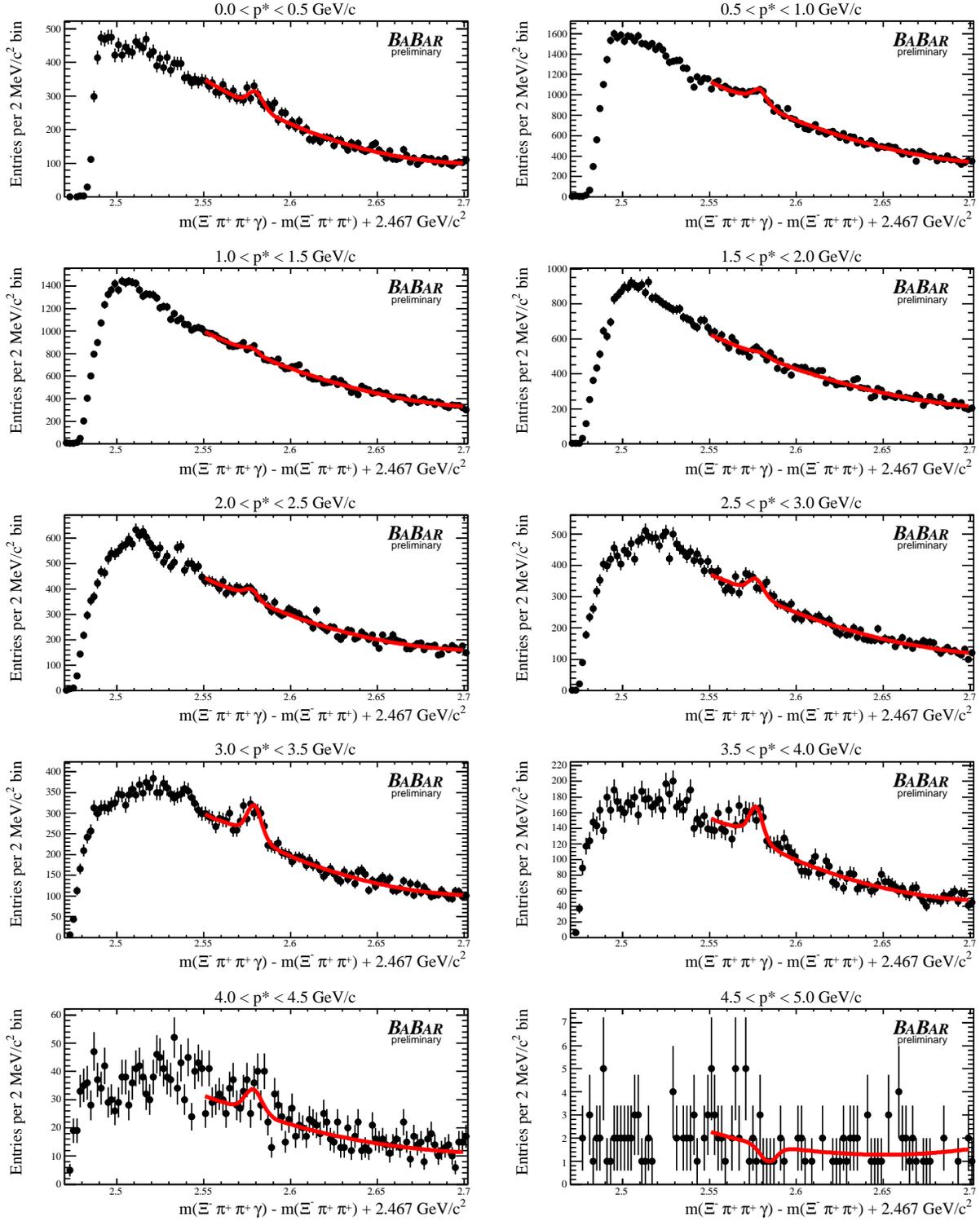, width=1.0\textwidth}
  \end{center}
  \caption{The $\Xi_c^+ \gamma$ invariant mass spectra for each of the
  ten $p^*$ intervals used. The points show data from the $\Xi_c^+$
  signal mass region (within $3\sigma$ of the central value).
  }
  \label{app:fig:xicplusprime}
\end{figure}

\begin{figure}
  \begin{center}
    \epsfig{file=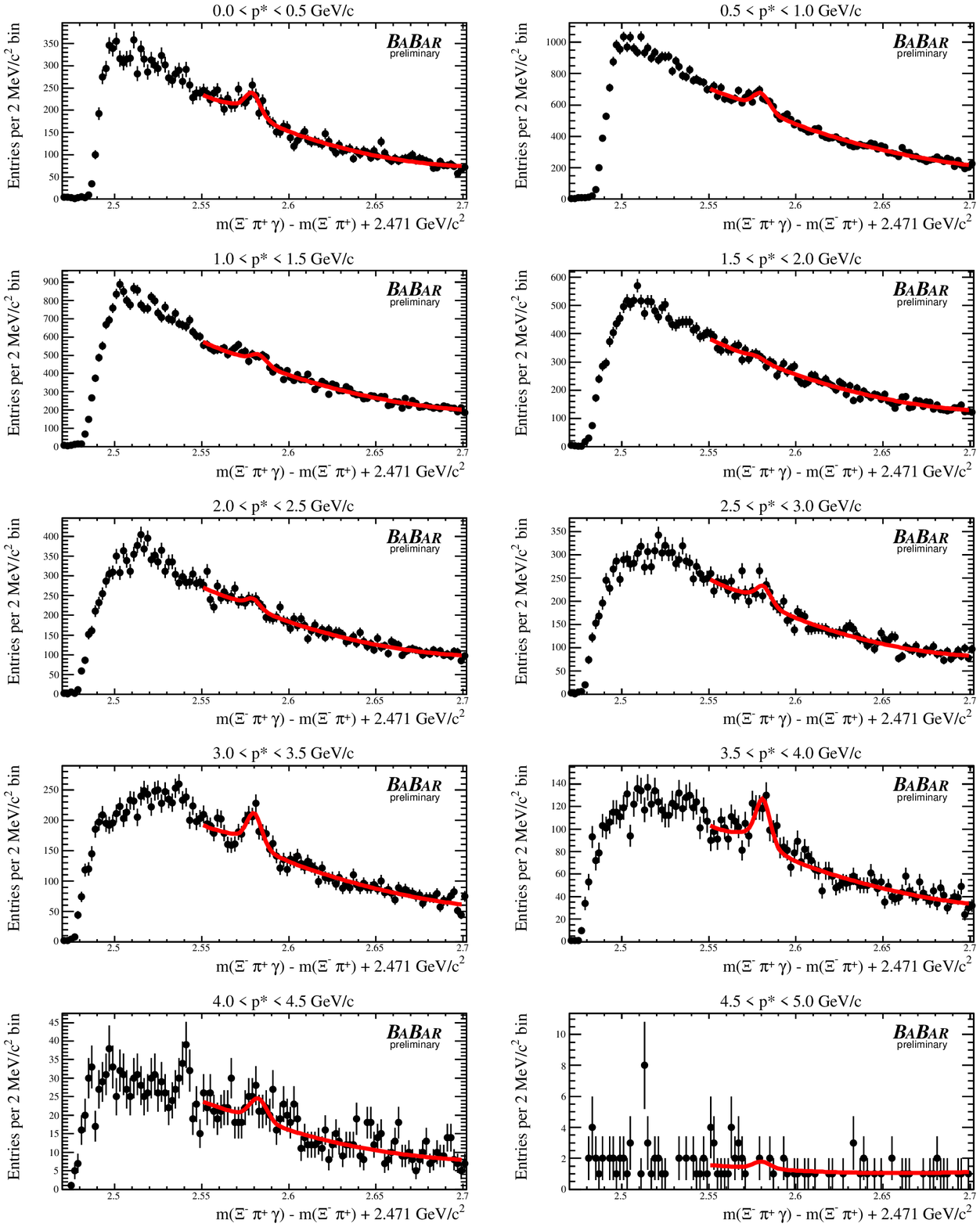, width=1.0\textwidth}
  \end{center}
  \caption{The $\Xi_c^0 \gamma$ invariant mass spectra for each of the
  ten $p^*$ intervals used. The points show data from the $\Xi_c^0$
  signal mass region (within $3\sigma$ of the central value).}
  \label{app:fig:xiczeroprime}
\end{figure}

The fitted mass spectra for individual $p^*$ intervals
are shown in Fig.~\ref{app:fig:xicplusprime}
and~\ref{app:fig:xiczeroprime}
for $\Xi_c^{'+}$ and $\Xi_c^{'0}$, respectively.

\end{document}